\documentclass[twocolumn]{openjournal}
\usepackage[dvipsnames]{xcolor} 
\usepackage{graphicx}
\usepackage{amsmath}
\usepackage{bm}
\usepackage[english]{babel}
\usepackage{natbib}
\usepackage{orcidlink}
\usepackage{microtype}
\usepackage{enumerate}
\usepackage[normalem]{ulem}

\newcommand{\msun}{\ensuremath{M_\odot}}

\begin{document}

\title{Differentiable Halo Mass Prediction and the Cosmology-Dependence\\ of Halo Mass Functions}

\author{\vspace{-1.5cm}Jim Buisman\,\orcidlink{0009-0008-9861-0582}$^{1,\star}$}
\author{Florian List\,\orcidlink{0000-0002-3741-179X}$^{1}$}
\author{Oliver Hahn\,\orcidlink{0000-0001-9440-1152}$^{1,2}$}

\affiliation{$^{1}$Department of Astrophysics, University of Vienna, Türkenschanzstraße 17, 1180 Vienna, Austria\\
$^{2}$Department of Mathematics, University of Vienna, Oskar-Morgenstern-Platz 1, 1090 Vienna, Austria}

\email{$^\star$Corresponding author: jim.buisman@univie.ac.at}

\begin{abstract}
\noindent Modern cosmological inference increasingly relies on differentiable models to enable efficient, gradient-based parameter estimation and uncertainty quantification. Here, we present a novel approach for predicting the abundance of dark matter haloes and their cosmology dependence using a differentiable, field-level neural network (NN) model, and study how well the cosmology dependence is captured by common parametrisations of the halo mass function (HMF), and by our NN-based approach. By training a 3D U-Net on initial density fields from fast $N$-body simulations with varying cosmological parameters, we enable direct, differentiable mapping from the linear density field to protohalo patches and their mass bins. Our method achieves competitive accuracy in identifying protohalo regions and in capturing the dependence of the HMF on cosmological parameters. Our NN derivatives agree well with finite differences of both analytical and emulated HMFs, at the level of the disagreement among the different models. We further demonstrate how the NN model can additionally be used to investigate the response of the HMF to changes in the initial Gaussian random field. Finally, we also demonstrate that a differentiable model can be used to extrapolate existing models at very high precision. 

\end{abstract}

\maketitle



\section{Introduction}
Dark matter haloes are the building blocks of structure in cold dark matter (CDM) cosmologies. Originating from the tiny density fluctuations thought to be seeded during the epoch of inflation \citep{KofmanLinde:1987}, they are the end product of hierarchical gravitational collapse, proceeding from small haloes merging over cosmic time to host the galaxy clusters of $10^{15}\,\msun$ in the low-redshift Universe. CDM haloes host the visible galaxies \citep[e.g.][for a review of the halo--galaxy connection]{Wechsler.2018.galaxies-and-dark-matter-haloes}, but their masses can extend to much smaller scales, depending on the properties of the dark matter particle -- and e.g.\ as low as earth mass if dark matter is made up from $100$ GeV WIMPs.

Due to its exponential suppression at the high-mass end -- probed by massive galaxy clusters today or by luminous galaxies at high redshift -- the halo mass function (HMF) has long been known as a sensitive probe of cosmology \citep[e.g.][]{Eke:1996}. The properties of the dark matter particle similarly impact the low-mass end \citep[e.g.][]{Angulo:2013}. Major efforts have gone into predicting its precise form. These predictions of the abundance (or number density) of collapsed structures date back to the pioneering work of \cite{Press-Schechter.1974.spherical-overdensity} based on thresholded (assuming spherical collapse) and filtered Gaussian random fields. Later refinements include the Extended-Press-Schechter (EPS) excursion set formalism of \cite{Bond.1991.excursion-set}, and the `peak patch' model of \cite{Bond.1996.peak-patch_halo-finder}. \cite{Sheth.2001.hmf} modified the spherical collapse model to reflect ellipsoidal collapse, further improving the agreement between theoretical predictions of the abundance of collapsed structures and simulations. \cite{MussoSheth:2012} and \cite{ParanjapeSheth:2012} included the peak constraint, and more recently \cite{MussoSheth:2021} found that peaks in energy rather than just density might be the most suitable field on which to carry out the excursion set random walks. A recent study by \citet{Wislocka.2025.perfect-collapse} showed that the assignment of a mass-scale to protohalo patches in the classical excursion set approach is plagued by large scatter (and bias) even if one can neglect errors from the collapse model.

In parallel to these theoretical developments, $N$-body simulations began to provide increasingly more accurate empirical measurements over increasingly larger mass ranges. Original Press-Schechter (PS) theory predicts universality, i.e.\ the HMF has a universal functional form when expressed in a suitable dimensionless variable $\nu$, and becomes independent of cosmic time and cosmology. In PS-theory the dimensionless variable is the peak height $\nu=\delta_c/\sigma(M)$, where $\delta_c$ is the critical threshold for collapse and $\sigma^2(M)$ is the variance of the linear density field filtered on a scale $R(M)=(3M/4\pi\overline{\rho}_m)^{1/3}$, given the mass $M$ of the structure and the mean matter density $\overline{\rho}_m$. The universal HMF then takes the form
\begin{equation} \label{eq:hmf}
    \frac{d n}{d \ln M} = f(\nu) \, \frac{\overline{\rho}_m}{M} \, \left|\frac{d \ln \sigma}{d \ln M}\right| \;,
\end{equation} 
where $f(\nu)$ is the so-called multiplicity function. Deviations from universality would be reflected in $f$ being also a function of cosmic time $t$ or of the cosmological parameters $\theta$, i.e.\ in full generality $f=f(\nu,\theta,t)$. Assuming universality, few $N$-body simulations suffice to fit this universal form very accurately \citep{Jenkins:2001,Reed:2003,Crocce:2010,Pillepich:2010,Angulo:2012,Watson:2013,Bocquet:2016} and allow extrapolations across masses and cosmic time. At the same time, on theoretical grounds, universality can only ever hold if the HMF depends on cosmology only through $\sigma(M)$ -- other quantities such as the growth rate 
$d \log D_+/d \log a$ of linear perturbations, or changes in the shape of the matter power spectrum that do not affect $\sigma(M)$ on the relevant scales will never enter. Also empirically, simulations found deviations from universality, to a degree depending on which halo definition is used, complicating accurate fits as a function of halo definition and cosmology \citep{Tinker.2008.hmf,Despali:2016,Diemer:2020,OndaroMallea:2022}. Most recently, large sets of high-resolution simulations covering relevant regions of cosmological parameter space and cosmic time have been used to train an `emulator' \citep{Bocquet.2020.mira-titan_HMF}, i.e.\ a non-parametric fit (or interpolation) of the mass functions measured from the simulations, to yield the most accurate predictions of the HMF yet. 

While conceptually a halo is usually thought of as a `virialised' spatially isolated overdense structure, in practice various heuristic approaches are used to define what is a halo \citep[cf.][for an overview]{Knebe.2013.halo-finders}. The most common approaches include spherical overdensity (SO) halo finders \citep{Efstathiou:1985}, which identify density peaks and collect particles in shells around them until the density drops below a chosen value (often based on a factor of the critical density of the universe, see also \citealt{Bryan.1998.halo-statistics}), and the `friends-of-friends' algorithm \citep{Davis.1985.FoF}, which connects particles within a specified linking length and does not assume spherical symmetry \citep[but see e.g.][]{Despali:2013}. Many more algorithms exist for the identification of subhaloes, i.e.\ self-bound substructures inside virialised main haloes -- we kindly refer the interested reader to \cite{Onions:2012} for a somewhat older overview and comparison. In this paper, we will however not be concerned with the identification of substructures.  

In parallel to the identification of virialised structures in non-linear simulations, there have always also existed approaches to predict collapsed structures directly from the linear density field -- thereby bypassing entirely the non-linear $N$-body simulation. The original peak-patch idea of \cite{Bond.1996.peak-patch_halo-finder} has been developed further for the precision era \citep{Stein.2019.halonet}, but more recently several groups have attempted to realise the same programme based on machine-learned models. Intuitively, the scale-space approach of EPS applied `at the field level' bears a fundamental similarity to what convolutional neural networks (CNNs) should be able to capture. \citet{Berger.2019.unet-halo-peak_patch} presented the first CNN-based method for predicting haloes, framed as a semantic segmentation problem (halo vs. non-halo particles), followed by a three-step algorithm for generating halo catalogues relying on connected regions in Lagrangian space. \cite{Bernardini.2020.unet-halo-edt} used NN based regression to assign a probability of halo membership of a Lagrangian fluid element in the linear density field, followed by a watershed step to group these fluid elements into contiguous Lagrangian protohalo patches. \citet{Lucie-Smith.2024.CNN-halo_mass} studied how well a CNN can predict the halo mass (and its possible secondary dependencies) from a given protohalo patch. \cite{Lopez.2024.instance-segmentation} used a two-step NN pipeline: first, a semantic segmentation network predicts which particles will collapse into haloes (binary prediction), and then an instance segmentation network groups these particles into individual haloes. This is achieved by mapping particles into a latent space where distinct haloes become more easily separable by a clustering algorithm, leading to final protohalo patches that are remarkably close to the ground truth. 

While these machine learning approaches can be viewed as powerful, modern (however arguably less interpretable) renditions of excursion set techniques, in that Lagrangian fluid elements are directly mapped to halo properties, they typically rely on non-differentiable algorithmic ingredients to construct discrete halo catalogues, such as watershedding or clustering, often applied in a post-processing step. In this regard, they share an important limitation with traditional simulation-based pipelines combined with FoF or SO halo finders, which also use operations (i.e., thresholding, grouping, and particle reassignment) that introduce discontinuities in the mapping from the linear density field to the final haloes. As a consequence, the HMF derived from such catalogues cannot be differentiated w.r.t.\ cosmological parameters or initial conditions, which makes it incompatible with gradient-based inference pipelines for cosmic large-scale structure observables.

In this paper, we are concerned with how HMFs can be construed that are \textit{differentiable} w.r.t.\ cosmological parameters, and even w.r.t.\ the specific realisation of the simulated universe. A key motivation for our aim is that differentiable mass functions can be employed in inference pipelines relying on gradient information, such as Hamiltonian Monte-Carlo \citep[HMC, cf.][]{Neal:2011:HMC}, which dramatically improves the sampling of posterior distributions in high-dimensional parameter spaces. In large-scale structure cosmology, HMC sampling has recently garnered significant attention as a central ingredient of \emph{field-level inference} (e.g.\ \citealt{2013MNRAS.432..894J}), whose promise is to provide more precise constraints on cosmological parameters than traditional analyses using summary statistics (e.g.\ \citealt{2024PhRvL.133v1006N}). Multiple differentiable forward models have been developed that are suitable for this task (e.g.\ \citealt{2021A&C....3700505M, 2022arXiv221109958L, 2025arXiv251005206L, 2025MNRAS.544..960H, 2025arXiv251213403H, 2025MNRAS.541..687K}), but they typically compute continuous field quantities and stop short of identifying haloes in view of the above-mentioned discrete components.\footnote{A promising approach that is orthogonal to this present work is the development of differentiable halo finders relying on soft halo membership assignments or stochastic gradients, e.g.\ via the REINFORCE estimator \citep{Horowitz.2025.jFoF}, see also \citep{2024MNRAS.529.2473H, 2026arXiv260204977P} for differentiable halo occupation distribution.}

For a universal HMF, or a parametric form of the multiplicity function $f(\nu,\theta)$ even in the non-universal case, obtaining gradients of the HMF is of course trivial if one can take derivatives of $\sigma(M)$ w.r.t.\ the cosmological parameters. This is possible with a differentiable Einstein-Boltzmann solver \citep{Hahn:2024:discoeb} or power-spectrum emulator \citep{Piras:2023:cosmopowerjax}, which we will investigate in this paper. Our main concern, however, is whether we can directly predict the abundance of collapsed structures in a given mass bin at the protohalo patch level with CNNs, and to what degree this prediction also captures cosmology dependence. Specifically, here we focus on two key-aspects relevant for cosmological inference:
\begin{enumerate}[(1)]
    \item the field-level mapping between a Lagrangian fluid element and its final halo mass (incl. whether it is collapsed to a halo at all)
    \item the dependence of this mapping -- and of the HMF itself -- on cosmological parameters, with particular emphasis on quantifying and differentiating the mapping and the HMF with respect to these parameters.
\end{enumerate}

In the context of field-level inference, it is usually necessary to compute gradients also w.r.t.\ to the initial density fluctuations, either in real or Fourier space. To showcase the feasibility of this within our framework, we also compute the HMF sensitivity to the initial density field amplitudes at fixed cosmology, which provides an intuitive picture of how different Lagrangian scales influence distinct mass ranges in the HMF.

The structure of this article is as follows. In the first part, we investigate how well we can predict the mapping between a Lagrangian fluid element and its $z=0$ final halo mass using a neural network (NN). Specifically, in \autoref{sec:nnmapping:methods}, we discuss the details of our chosen NN architecture and the training/validation dataset, while in \autoref{sec:nn:validation}, we present an analysis of the performance, especially in capturing the cosmology dependence.

\section{Field-level predictions of halo mass -- Methodology}
\label{sec:nnmapping:methods}
In this section, we present our methodology for predicting halo masses at the field level from the initial linear density field using a CNN. We first describe the generation of training and validation data using fast $N$-body simulations in \autoref{sec:nbody}. Next, we detail the NN architecture and training procedure in \autoref{sec:NN}. Finally, we assess the performance of our approach and validate the results in \autoref{sec:nn:validation}.

\subsection{Data generation using fast N-body simulations}
\label{sec:nbody}
Since our main task will be to assign a halo mass to all Lagrangian fluid elements using a CNN under varying cosmological parameters, a sufficiently large training/validation dataset is required. We use fast $N$-body simulations run with the \textsc{Disco-Dj} code \citep{List.2025.disco-dj-PREP} for this purpose. The full training+validation set consists of 2000 independent simulations before data augmentation. All results presented herein are based on independent test simulations.

\paragraph{Cosmological model} We adopt a flat $\Lambda$CDM cosmology and vary the matter density parameter $\Omega_m$, the linear power spectrum normalisation as quantified by $\sigma_8$, and the primordial scalar spectral index $n_s$. We keep all other parameters fixed at their fiducial values; specifically we choose a dimensionless Hubble constant $h=0.677$ and a baryon density parameter of $\Omega_b=0.049$. Due to the flatness assumption, the dark energy density is related to the matter density as $\Omega_\Lambda = 1 - \Omega_m$ and is therefore not treated as an independent parameter. Importantly, this means that derivatives w.r.t.\ increasing $\Omega_m$ correspond to decreasing $\Omega_\Lambda$ and vice versa. The varying set of parameters is sampled from a three-dimensional cube using a Latin hypercube sampling scheme \citep{McKay.1979.latin-hypercube}, covering the range shown in \autoref{tab:CNN_cosmo-parameter-limits}. For simplicity, we adopt the parametrisation of \citet{Eisenstein1998BaryonicFunction} for the matter transfer function, but we checked that computing the transfer function with an Einstein-Boltzmann solver only makes a small difference ($< 6\%$) for the HMF and its gradient w.r.t.\ cosmological parameters. We note that our NN is not conditioned on the cosmological parameters (i.e.\ they are not directly provided
as input). This is important because it forces the NN to learn everything from the initial density field, which itself depends on the cosmology through the power spectrum. In this way, the model is not limited to any specific set of parameters, but it will of course not be sensitive to any parameters that affect only the growth $d D_+/d a$. Furthermore, it can be used without any prior knowledge of the parameters. The chosen upper and lower bounds for each parameter are based on the Quijote simulation suite \citep{2020.Quijote_sims}, as well as the \textsc{Mira-Titan} HMF emulator \citep{Bocquet.2020.mira-titan_HMF} to cover the entire range of their parameter space. 

\begin{table}
\centering
\caption{\normalfont \it Cosmological parameter limits for the Latin hypercube distribution of training simulations.}
\begin{tabular}{lccc}
\hline
 & $\Omega_m$ & $n_s$ & $\sigma_8$ \\
\hline
Lower Bound & 0.16 & 0.85 & 0.7 \\
Upper Bound & 0.52 & 1.05 & 0.9 \\
\hline
\end{tabular}
\label{tab:CNN_cosmo-parameter-limits}
\end{table}

\paragraph{Non-linear model} We perform GPU-accelerated particle-mesh (PM) simulations with the \textsc{Jax}-based \citep{jax.2018.github} \textsc{Disco-Dj} framework, using $N = 256^3$ particles. We employ a PM grid with $M = 1024^3$ cells, and the simulation is evolved to $z=0$ in 100 time steps with the recently presented \textsc{BullFrog} time integrator \citep{Rampf.2024.bullfrog, List2023Perturbation-theorySimulations}. On mildly non-linear scales, the knowledge of LPT built into \textsc{BullFrog} enables accurate results with significantly fewer steps than for a standard leapfrog integrator. The comparably high grid resolution in comparison to the number of particles is chosen in order to balance accuracy and computational efficiency. The inherent underestimation of small-scale forces by the PM method -- which could potentially give rise to less massive haloes -- is significantly mitigated by the high grid resolution. On the other hand, the lower particle resolution keeps the computational costs in terms of runtime and memory manageable, particularly for the CNN training. One simulation could thus be run in $\sim 20\,\mathrm{s}$ on an NVIDIA A100 GPU.

\paragraph{Halo finding} We use \textsc{Rockstar} \citep{Behroozi.2013.Rockstar} to identify all dark matter haloes and exclude all subhaloes to ensure that each particle is unambiguously assigned to either the background or an individual halo. We employ the spherical overdensity criterion of \cite{Bryan.1998.halo-statistics} to define SO haloes as our targets, since many analytical functions are based on these mass definitions and they are observationally better motivated \citep{Tinker.2008.hmf}. Given that \textsc{Rockstar} only saves the FoF particle IDs, we extracted the SO halo particles in an additional post-processing step of the \textsc{Rockstar} output. In case of overlapping haloes, the particles are counted towards the closer centre.

For a given resolution, increasing the box size improves halo statistics and reduces the impact of cosmic variance, which is especially important for rare, massive haloes at the high-mass end. However, this comes at the cost of lower resolution at the low-mass end. As a trade-off, we choose a volume of $V_{\text{box}}=(100 \text{ Mpc}/h)^3$. This means that on average, every second simulation box contains a halo of mass $10^{15} M_{\odot}/h$. Similarly, there are about $600$ haloes of mass $4 \times 10^{12} M_{\odot}/h$, with each containing approximately $800$ particles. 

\begin{figure}
    \centering
    \includegraphics[width=\linewidth]{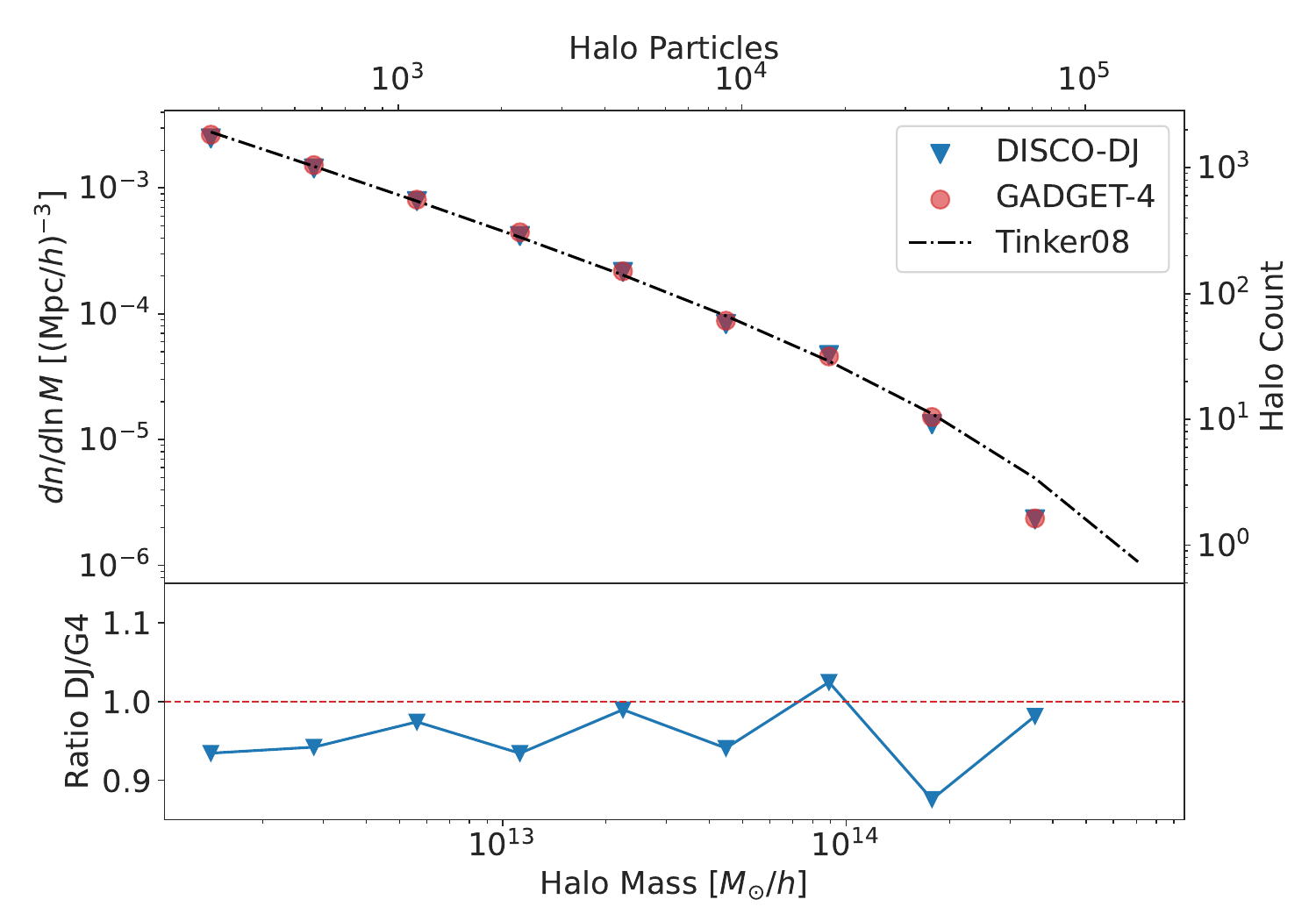}
    \caption{Halo mass function (HMF) comparison of \textsc{Disco-Dj} (blue triangles) and \textsc{Gadget-4} (red dots) for a single simulation with the same initial conditions and particles ($256^3$). The top graph shows the two $N$-body simulators compared to the analytical Tinker08 HMF (dashed line). The bottom shows the difference between both codes with \textsc{Gadget-4} being used as a reference line.}
    \label{fig:hmf_dj_vs_g4}
\end{figure}

\paragraph{Validation \& convergence} In \autoref{fig:hmf_dj_vs_g4}, we show the HMF obtained from a single \textsc{Disco-Dj} run and compare it to a \textsc{Gadget-4} \citep{Springel:2021:Gadget4} simulation with the same initial conditions, as well as the HMF predicted by the \cite{Tinker.2008.hmf} (Tinker08) parametrisation for this cosmology. We find that our simulation accurately reproduces the HMF with an average deviation of $< 5\%$ over the full mass range considered in this work. The largest halo mass bin deviates slightly from the predicted Tinker08 line, which is simply due to cosmic variance and low number counts (see the right $y$-axis, which indicates the total halo count in the box). In \autoref{sec:app:convergence_nbody}, we provide additional material demonstrating the dependence (and convergence) of the measured HMF as a function of PM resolution. Our finding that the HMF based on fast simulations is biased low is in consistent with the results of \citet{Wu:2024:fastpm_hmf} for \textsc{FastPM} \citep{Feng2016FastPM:Haloes}. These authors also found that by adjusting the overdensity at low particle numbers the HMF can easily be corrected, a strategy that we however decided to not follow here. 

\subsubsection{Differentiable halo field responses}

A central goal of this work is not only to predict the halo field itself but also to capture how it responds to perturbations of cosmological parameters and the initial conditions. To this end, we formulate halo finding as a fully differentiable field-level prediction problem.

\begin{figure*}
    \centering
    \includegraphics[width=\linewidth]{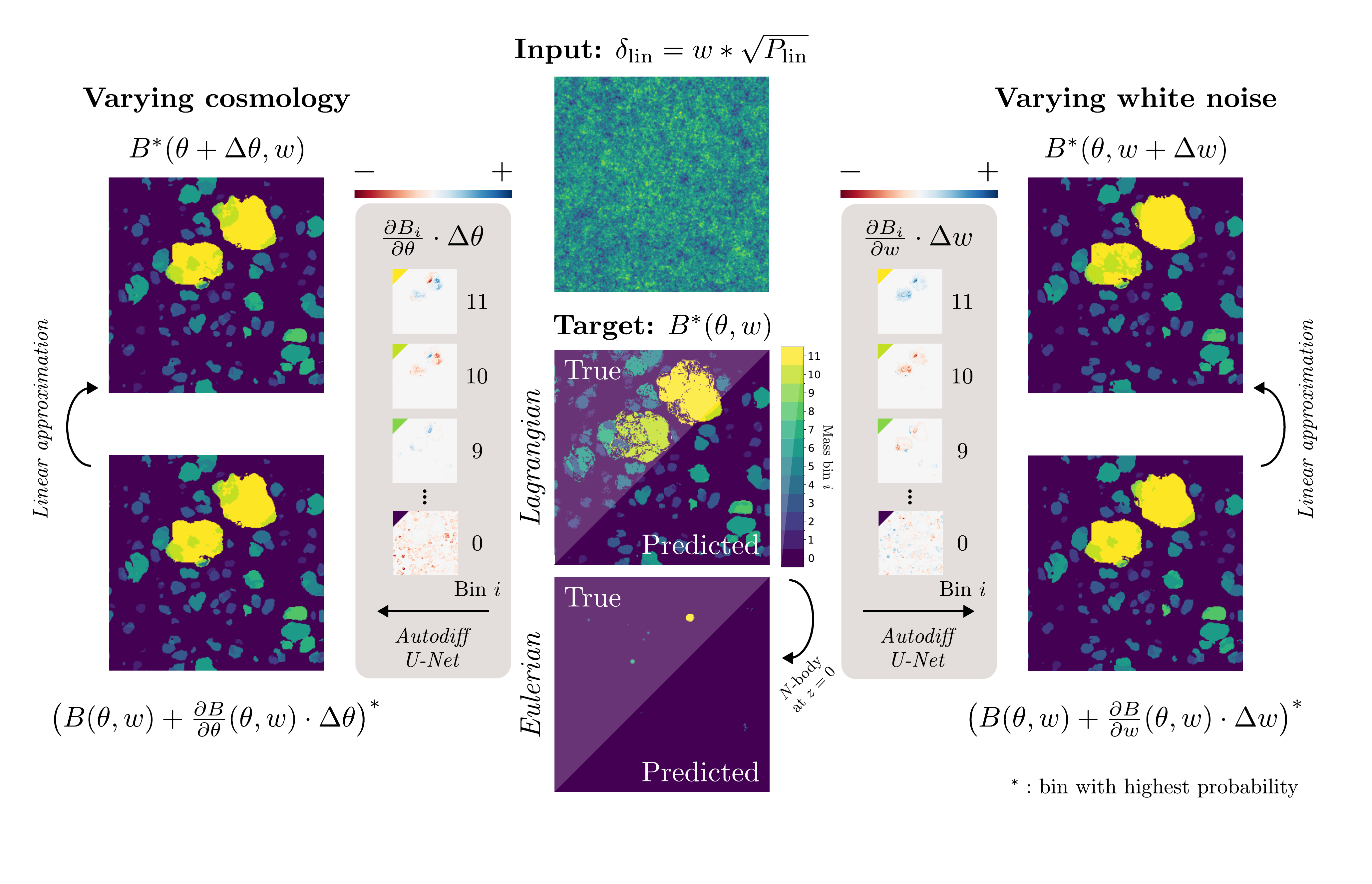}
    \caption{This figure illustrates the different variations on the linear density field and their effect on protohalo patches that we will analyse throughout the paper. All plots show the same slice of a three-dimensional simulation box ($256^3$ particles) with a length of $100 \text{ Mpc}/h$. The linear density field (central top) is used as input data for the NN. The model tries to predict which Lagrangian fluid elements will end up in haloes at $z=0$ and probabilistically assign them to one of 12 mass bins (central middle panel, diagonally split between the ground truth as determined by \textsc{Rockstar} and the NN prediction). The background is represented by class 0. The target is given by the mass bins of the Lagrangian protohalo patches. The central bottom panel plots the same physical field, but in Eulerian coordinates at $z = 0$. A key objective of this work is to study how protohalo fields respond to perturbations of cosmological parameters and the initial white noise. The top left panel shows the  halo predictions by the NN in Lagrangian space for a slightly perturbed cosmology. At linear order, the response for every mass bin can be captured using gradient information, computed with automatic differentiation (left grey). Using a bin-wise Taylor correction, one obtains a linearisation of the perturbed Lagrangian halo patches (bottom left). The exact same idea can also be applied to variations in the white noise (right side).}
    \label{fig:full_overview}
\end{figure*}

\autoref{fig:full_overview} illustrates the overall framework used throughout this paper. The linear density field is used as input to the U-Net, which predicts protohalo patches in Lagrangian space and probabilistically assigns them to one of 12 mass bins. The central middle panel compares the ground truth protohalo patches obtained from \textsc{Rockstar} with the corresponding NN prediction. The background is represented by class 0. Since the entire pipeline is differentiable, automatic differentiation can be used to quantify how the predicted halo field changes under perturbations of either the cosmological parameters or the initial white noise field. The upper left and right panels show the corresponding perturbed halo predictions in Lagrangian space. The grey panels illustrate the local gradient response for individual halo mass bins. These gradients further allow us to construct first-order Taylor approximations of the perturbed halo fields, shown in the lower panels. This provides an efficient way to estimate the response of the halo field without explicitly reevaluating the network for every perturbation, and allows integrating our NN into differentiable inference pipelines. Throughout this work, we primarily analyse halo fields in Lagrangian space, where protohalo patches remain spatially extended and individual structures are visually distinguishable. For comparison, the central bottom panel shows the same halo field in Eulerian space at $z=0$, where gravitational collapse and the visualisation in comoving coordinates significantly compresses the spatial extent occupied by individual haloes.

\paragraph{CNN input data}
The input to the U-Net is the linear density field, normalised by the mean standard deviation of the linear overdensity at $z = 0$ across the entire dataset, which accelerates convergence during training. The target labels correspond to mass-binned protohalo patches in Lagrangian space. We use a total of 12 classes. The inner 10 bins contain halo masses between $4 \times 10^{12} M_{\odot}/h$ and $1 \times 10^{15} M_{\odot}/h$ with logarithmic spacing. Lagrangian fluid elements corresponding to masses below the minimum threshold are labelled as background (bin 0) and are not used for calculating the HMF. At the high-mass end, we use the uppermost bin to collect all haloes above $1 \times 10^{15} M_{\odot}/h$ during the NN training. However, since this bin lacks a natural upper edge, we cannot assign it a well-defined mean mass -- which is required for computing the HMF -- and therefore exclude it from our statistical analyses. As a result, the network predicts 10 effective mass bins for the protohalo patches.

\subsection{NN architecture \& training}
\label{sec:NN}
We formulate halo finding as a multiclass semantic classification task where each Lagrangian fluid element (voxel) of the initial linear density field is to be classified as belonging to a given mass bin or to the background. We use the notions of Lagrangian fluid elements, voxels, and particles (each of which is associated with a Lagrangian fluid element in the $N$-body simulation) interchangeably, depending on the context.

\begin{figure*}
    \centering
    \includegraphics[width=\linewidth]{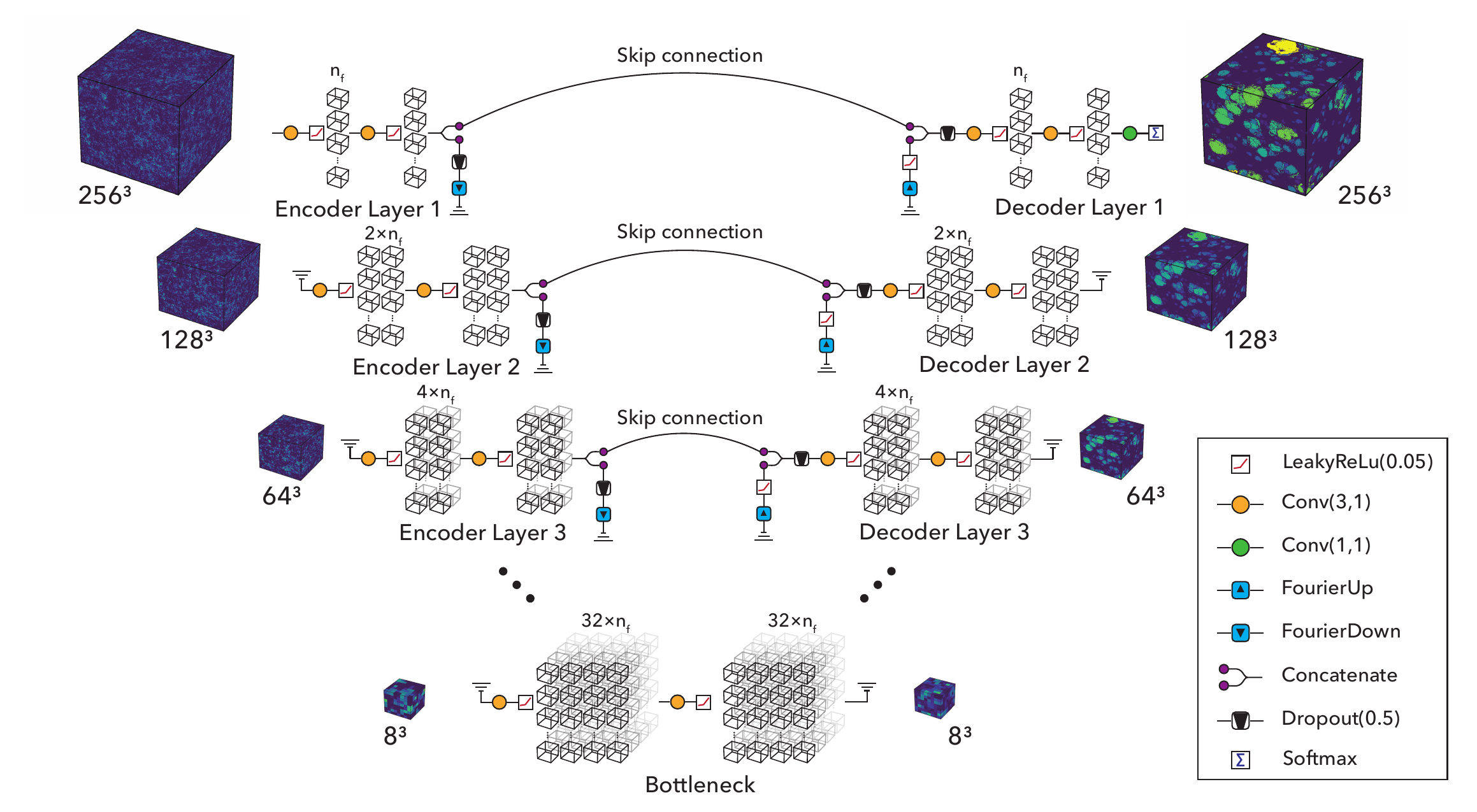}
    \caption{A schematic representation of the U-Net architecture of depth 6 used for classification. The legend shows each module and corresponding parameters used, i.e.\ (kernel size, stride) for the convolutions, the slope for negative inputs for LeakyReLU, and the dropout probability. The number of filters of the first convolution is $n_f=16$, changing with each following layer. The final convolution uses 12 filters, which represent 10 (excluding background and haloes above the maximum mass) effective mass bins.}
    \label{fig:cnn_architecture}
\end{figure*}

\paragraph{Architecture}  We use a U-Net (also known as V-Net in 3D) architecture \citep{ronneberger2015u}, implemented using the \textsc{Flax} \citep{flax.2020.github} NN library for \textsc{Jax}. It consists of three parts: the encoder (left side), bottleneck (bottom middle), and decoder (right side), as can be seen in \autoref{fig:cnn_architecture}. 

The encoder is composed of consecutive down-sampling blocks that allow the CNN to assess the density fields at multiple spatial resolutions. As the spatial resolution decreases with each layer, the number of channels increases, forcing the network to compress spatial features into a more abstract representation. The maximum compression occurs at the bottleneck: at this stage, the input has been down-sampled from $256^3$ to $8^3$ voxels, allowing a single convolution kernel (of size $3\times3\times3$) to analyse a volume of $(37.5 \text{ Mpc}/h)^3$. At the same time, the number of channels has grown to $512$. 

The decoder then reconstructs the spatial structure of the protohalo patches from these compressed representations of the linear density fields. At each resolution level, intermediate outputs from the encoder are passed to the decoder via `skip connections', which help retain spatial detail lost during down-sampling. After the final decoder layer, the output has been restored to the original resolution of $256^3$, with $12$ channels (corresponding to the mass bins). Finally, a softmax function is used to obtain probabilities that add up to unity over the classes.

The layers of the encoder and decoder use standard building blocks such as convolutions (for feature extraction), a `Leaky-ReLU' activation function (introducing non-linearity), and dropout (to mitigate overfitting, see \citealt{Hinton.2012.dropout}). In view of the periodicity of the simulation box, we use periodic padding for all convolutions. While blockwise pooling (such as $\operatorname{MaxPool}$ over $2 \times 2 \times 2$ blocks) for down-sampling and transposed convolutions / bilinear interpolation for up-sampling are standard choices, we instead perform these operations in Fourier space (see also \citealt{Floess.2024.fourier}). Specifically, down-sampling is applied via a sharp-$k$ low-pass filter that removes all frequencies exceeding the 1D Nyquist limit at the down-scaled resolution. 
Up-sampling, in turn, is implemented by zero-padding the Fourier modes between the current and the next higher resolution grid. This ensures that the low-frequency content of the data remains unaltered by the resolution change, which is beneficial in cosmology, given that common metrics of interest such as the power spectrum are computed in Fourier space. 

\paragraph{Loss function} We use Adam gradient descent \citep{Kingma2014} with a learning rate of $10^{-5}$, which proved to be optimal for this setup when testing different values. Since the classes have natural ordering, the often used cross-entropy loss is not optimal since it does not take any relation between classes into account. Instead, we use the squared Earth Mover's Distance (EMD, see \citealt{Rubner.2000.EMD, Hou.2016.EMD-DNN}), which measures the minimum work required for transforming one probability distribution into another. Here, `work' refers to probability mass times distance, where the distance between each pair of classes is encoded by a distance matrix $\mathbf{D} = (D_{ij})$. In the case of ordered-class classification, such as in our application where each class corresponds to a halo mass bin, the distance matrix admits a one-dimensional embedding as $D_{ij} = |i - j|$. In that case, it can be shown \citep{Levina:2001:EMD} that the $\ell^q$ generalisation of the EMD can be computed straightforwardly by harnessing cumulative distribution functions (CDFs). More specifically, our loss function is
\begin{equation}
    L_{\mathrm{EMD}^q}(\bm{p}, \bm{t}) = \frac{1}{N^3} \sum_{n=1}^{N^3} \sum_{i=1}^{C} |\text{CDF}_i(p_n) - \text{CDF}_i(t_n)|^q \;,
    \label{eq:emd}
\end{equation} 
where $C$ are the classes, and $\bm{p} =  (p_n)_{n=1}^{N^3}$ and $\bm t = (t_n)_{n=1}^{N^3}$ the prediction and target values, respectively, in all voxels $n = 1, \ldots, N^3$. Note that we left out an unimportant normalisation constant. As proposed by \citet{Hou.2016.EMD-DNN}, we use $q = 2$, i.e.\ the EMD$^2$ loss. 
Intuitively, the EMD loss implies that the NN is penalised in proportion to the number of bins by which its prediction for a given input voxel differs from the true mass bin label. In particular, this loss function implicitly assumes that the ground distance between the background (non-halo) bin and the least massive halo bin is the same as the distance between any two adjacent halo bins. It would also have been possible to weigh the loss function in a way to increase the difference between background (class 0) and lightest mass bin (class 1), or to devise a more complex ground-distance definition altogether. However, we choose not to impose an artificial weighting scheme that might bias the classifier toward certain mass bins. Instead, we rely on the model to learn the natural distribution of the haloes without explicitly enforcing larger separations between specific classes. 

Note that if the true class label is $i_0$, one has $\text{CDF}_i(t) = 0$ for $i < i_0$ and $\text{CDF}_i(t) = 1$ for $i \geq i_0$ (omitting the voxel index for brevity), whereas the predicted values $\text{CDF}_i(p)$ are generally fractional numbers in $(0, 1]$ that increase monotonically up to $\text{CDF}_C(p) = 1$, as enforced by the softmax function through which $p$ is computed. In other words, while each voxel is assigned a single `hard' class label during training, the U-Net predicts a smooth, probabilistic estimate over halo mass classes -- including the probability that the voxel does not correspond to a halo at all.

\paragraph{Training \& data augmentation}
We split our sample of 2000 simulations into a training and validation dataset. More specifically, 1800 are used for updating the weights, while the other 200 are used as validation for hyperparameters and against overfitting. We perform additional data augmentation by rotating and flipping along each axis. With these augmentations, each simulation can be represented in $48$ unique ways, which leads to a total number of 96,000 data cubes available for training and validation. During each epoch, when a new batch of training data is loaded, the network randomly rotates and flips the data, while keeping input and target consistent. The NN is trained indefinitely until the validation loss has stopped improving over the last 100 epochs, at which point it is stopped automatically. This ensures that the model has had enough time to extract as many features as possible. The final model has been trained for 239 epochs. The entire process was performed on two NVIDIA A100 GPUs and took just under 40 hours.

\newpage
\section{Field-level predictions of halo mass -- Results and validation}
\label{sec:nn:validation}

In this section we apply the model from \autoref{sec:nnmapping:methods} to predict the $z=0$ halo mass for Lagrangian fluid elements. Specifically, we discuss the accuracy of the predicted HMF in \autoref{sec:results_hmf}.

\subsection{Labelling protohalo patches}
\label{sec:results_prediction}

\paragraph{Binary classification: halo or not halo?} 
Although our NN approach directly estimates halo masses, with non-halo voxels gathered in the lowermost mass bin (i.e.\ without a `binary classification detour' for making the decision `halo' vs. `no halo'), we first evaluate its performance by means of binary metrics that merely assess whether Lagrangian fluid elements have been correctly assigned to a protohalo or the background, neglecting halo masses.
Based on the fraction TP (TN) of voxels correctly classified as haloes (background), and the fraction FP (FN) of voxels incorrectly classified as haloes (background), we determine the usual classification metrics: 
\begin{center}
\begin{tabular}{ll}
    true positive rate & $\text{TPR} = \frac{\text{TP}}{\text{TP} + \text{FN}}$ \\[.1cm]
    true negative rate & $\text{TNR} = \frac{\text{TN}}{\text{TN} + \text{FP}}$ \\[.1cm]
    positive predictive value & $\text{PPV} = \frac{\text{TP}}{\text{TP} + \text{FP}}$ \\[.1cm]
    accuracy & $\text{ACC} = \frac{\text{TP} + \text{TN}}{\text{TP} + \text{TN} + \text{FN} + \text{FP}}$ \\[.1cm] 
    F$_1$-score & $\text{F}_1 = \frac{2\text{TP}}{2\text{TP} + \text{FP} + \text{FN}}$
\end{tabular}
\end{center}

We count a voxel as predicted to be in a protohalo if the cumulative probability assigned to non-background bins is at least 50\%.

We compare the values for our method (averaged over 100 predictions) with those given by \cite{Lopez.2024.instance-segmentation} for their method in \autoref{tab:performance_comparison}. For context, ExSHalos \citet{Voivodic.2019.ExSHalos} represents a deterministic approach based on the excursion set theory, while their Optimal values correspond to the upper limit of segmentation consistency, estimated by comparing the agreement between two $N$-body simulations with slightly different initial conditions.

\begin{table}[b] 
    \centering
    \caption{\normalfont \it  Performance metrics of our NN using only binary distinction to compare against the results from \citet{Lopez.2024.instance-segmentation}, which also contain ExSHalos, their own semantic segmentation model, and their optimal target accuracy estimated from the baseline simulations. The table presents True Positive Rate (TPR), True Negative Rate (TNR), Positive Predictive Value (PPV), Accuracy (ACC), and F$_1$-score.}
    \begin{tabular}{lccccc}
        \hline
        Type & TPR & TNR & PPV & ACC & F$_1$ \\
        \hline
        ExSHalos & 0.518 & 0.845 & 0.707 & 0.708 & 0.598 \\
        López-Cano & 0.838 & 0.883 & 0.838 & 0.864 & 0.838 \\
        Optimal & 0.887 & 0.914 & 0.882 & 0.903 & 0.884 \\
        \hline
        our U-Net & 0.732 & 0.956 & 0.822 & 0.907 &  0.774 \\
        \hline
    \end{tabular}
    \label{tab:performance_comparison}
\end{table}

Comparing the TPR and TNR of our model with those of López-Cano we see that our U-Net is rather conservative in predicting protohaloes.
Our NN achieves only 73\% TPR, while the TNR is extremely high with 96\%. This shows that our CNN approach is very accurate in predicting which particles will not end up in a halo. The overall accuracy of the U-Net is also very high with just below 91\%, which is better than the results of the other papers. However, our model is slightly worse than that of López-Cano et al. in terms of the F$_1$-score, which is more representative with unbalanced datasets, but still performs better than ExSHalos. Nonetheless, our classification network proves to be a fairly accurate and competitive model for identifying protohalo patches from the initial density field. 

\begin{figure*}
    \centering
    \includegraphics[width=\linewidth]{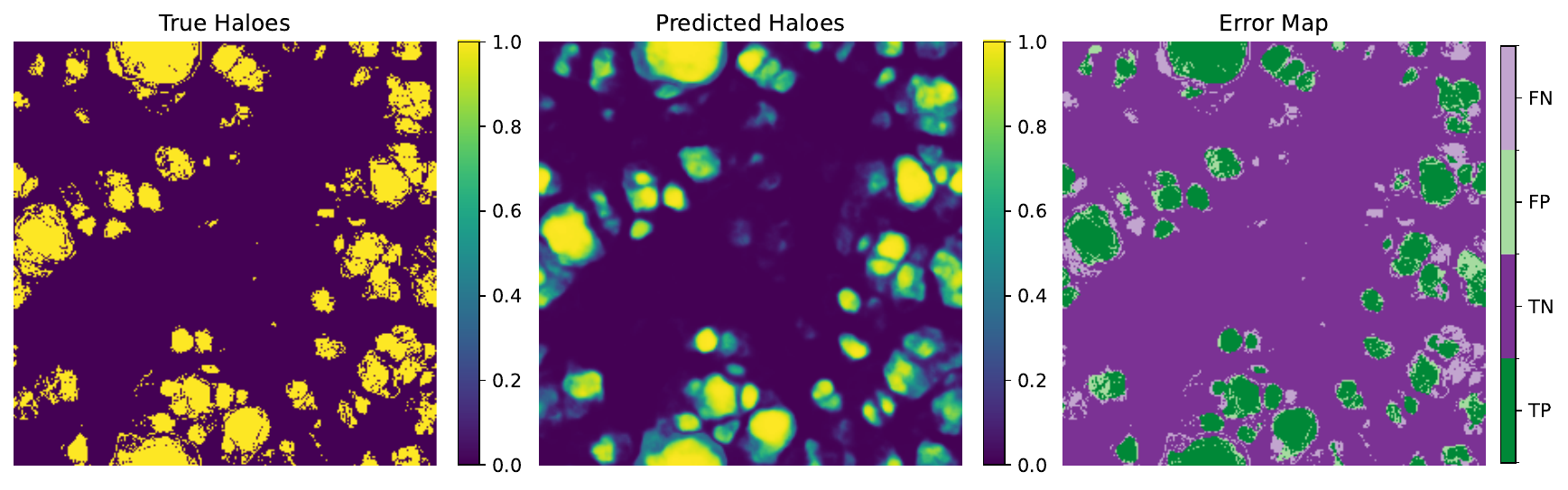}
    \caption{A thin slice of initial conditions showing how well the U-Net is able to identify protohalo patches on a voxel level for a test simulation. The left image shows the ground truth as identified by evolving a simulation with \textsc{Disco-Dj} to $z=0$, using \textsc{Rockstar} to find all haloes above a mass of $4 \times 10^{12} M_{\odot}/h$, and tracing the individual particle positions back to the initial conditions. The dark purple voxels represent background, while yellow regions will eventually end up in haloes. The middle image shows the prediction of the NN for the same input. The multiclass probabilities are aggregated to a binary classification showing the probability for each particle to land in a halo. The right image shows the consistency between both maps, where the maximum probability is used for both classes (halo vs. no halo) in the prediction. The colours represent True Positive (TP), True Negative (TN), False Positive (FP), and False Negative (FN).}
    \label{fig:errormap}
\end{figure*}

In order to better understand the accuracy of our model and identify the challenging regions, \autoref{fig:errormap} shows a visual representation of the correct and incorrect classifications within a slice of a single simulation. The ground truth (left) plot is simply a binary map in Lagrangian space, while the prediction (middle) contains the probability for each voxel to end up in a halo (of any mass). It is clear that the network is most certain about particles in the middle of a patch and less certain about the edges. This also applies to smaller haloes, where even the central particles are relatively close to background particles, making it harder for the network to distinguish them confidently. This uncertainty is expected due to the ambiguity of the haloes themselves, as well as the halo finder. The effect is also much stronger at lower resolutions, as we observed in previous experiments with our setup. 

However, we do not have to rely on just the most likely class. Instead, we can utilise the probabilities directly for statistical analysis, such as the HMF computation. We can clearly see a very small protohalo slice in the very centre of the image that does not have a single particle identified above $50\%$ confidence. However, compared to the ground truth, the model predicts a more extended patch at around $40\%$, which roughly adds to the same mass in this case.

To simplify the visual assessment of the model accuracy, the error map (right) is produced by converting the probabilistic predictions into a binary map using the maximum probability for both classes (i.e.\ halo or no halo). Dark green and dark purple represent correct predictions of protohalo patches (True Positive) and background (True Negative), respectively. Additionally, light green shows the model predicting halo particles when it should be background (False Positive), while light purple indicates missing protohalo voxels (False Negative). This shows that the majority of False Negatives lie along the edges of protohalo patches, where the assigned halo probability drops below the 50\% threshold. The False Positives appear most often in regions where the true patches are discontinuous or have more complicated shapes. The U-Net struggles with jagged edges or individual voxels inside of an protohalo patch that should not be part of the halo according to the $N$-body simulation and halo finder. Instead, the NN smooths the regions out, which causes more rounded and usually completely filled patches. Clearly, the labels should not be regarded as the `absolute truth' on all scales, given that the chaotic gravitational dynamics on halo scales might sensitively depend on numerical effects, such as the discrete sampling of the underlying continuous field by the discrete $N$-body particles, the limited machine precision, etc. As such, a smoother prediction might, in some cases, arguably be a more faithful representation of the (unobservable) truth than the actual label. This indetermination, which inherently limits the degree to which agreement between the prediction and the truth can be expected even with a perfect model, has been studied in more detail by \citet[e.g.\ Fig.\ 3]{Lopez.2024.instance-segmentation}.

\begin{figure*}
    \centering
    \includegraphics[width=\linewidth]{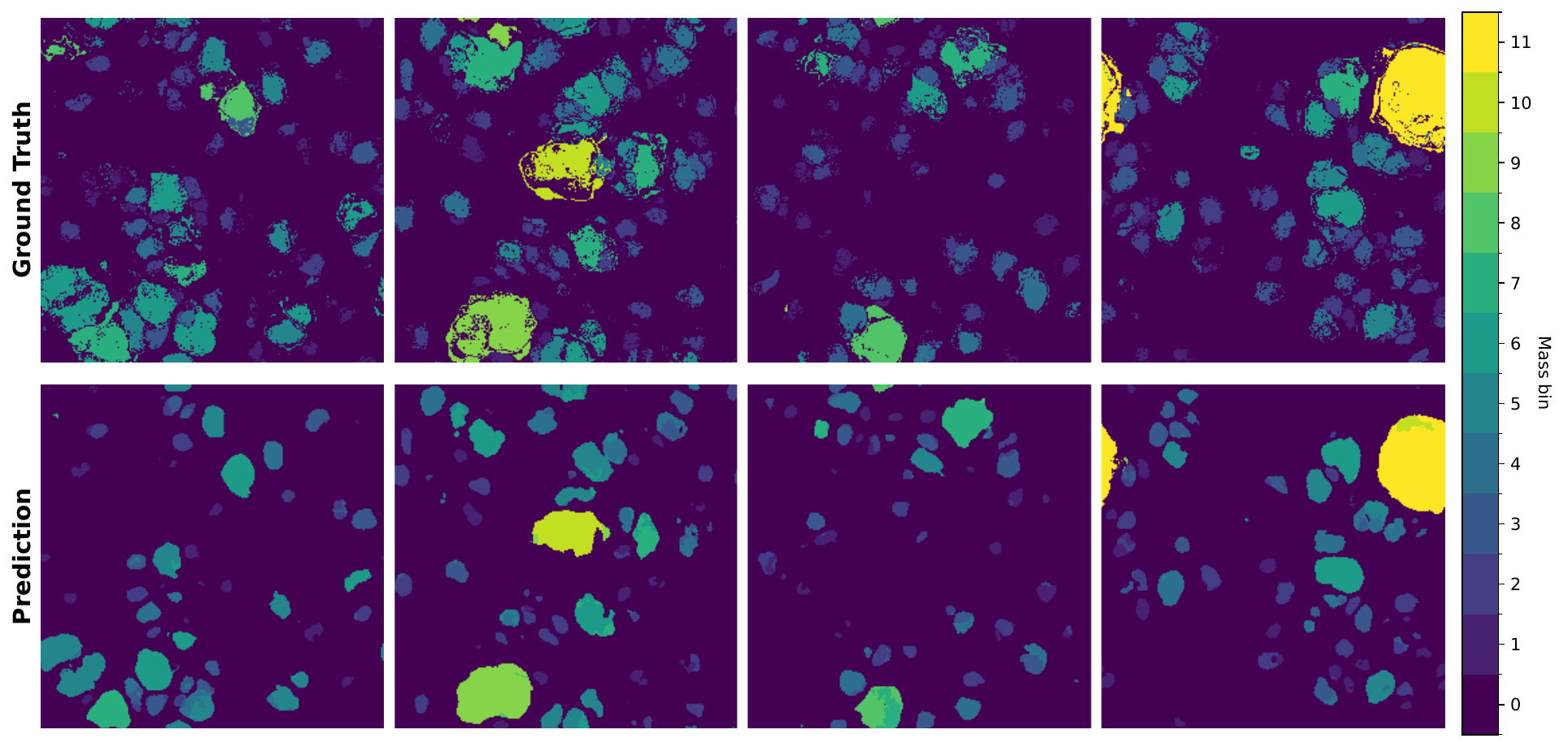}
    \caption{Random slices of the protohalo patches for different simulations in the test dataset. The upper row corresponds to the ground truth computed by $N$-body and \textsc{Rockstar}. The lower row shows the model predictions of the same slices and initial density fields (i.e.\ the mass bin to which the highest probability has been assigned). The colours represent the final virial mass of each halo at redshift $z=0$, logarithmically binned between $4 \times 10^{12} M_{\odot}/h$ and $1 \times 10^{15} M_{\odot}/h$. }
    \label{fig:multiple_predictions}
\end{figure*}

\paragraph{Multiclass: predicting halo mass bins} We next assess how well the network assigns halo mass bins to each Lagrangian fluid element (voxel). In \autoref{fig:multiple_predictions}, we show the model's multiclass output, where each protohalo patch is labelled by its predicted mass bin. This enables distinguishing adjacent haloes unless they share the same mass bin, in which case no boundary between the haloes can be drawn (as is always the case when performing semantic segmentation). The figure displays random slices from different test simulations, with each voxel coloured by the most probable class. Compared to the ground truth, the predictions are smoother, with small haloes sometimes missed or underestimated in size. Notably, the U-Net typically assigns a consistent mass bin across most of a halo, even as the probability drops near patch edges, avoiding spurious low-mass predictions around larger haloes. This robustness is particularly remarkable given the bin ordering imposed by the EMD loss, where low-mass bins lie between the background and high-mass bins. One might therefore worry about the model interpolating by assigning halo boundary regions to low-mass bins; yet, such spurious low-mass predictions are largely avoided. 
We present a theoretical argument for this using a toy example in \autoref{sec:app:emd-toy}.
Occasional misclassification at patch boundaries, especially for the most massive haloes, are rare and may reflect true neighbouring smaller haloes. Overall, the model reliably predicts mass-binned protohalo patches at the voxel level, with only minor smoothing and mass assignment differences. We also show a confusion matrix as a quantitative assessment of the multiclass performance in \autoref{fig:confusion_matrix} in \autoref{sec:app:confusion}. 
We next turn to a statistical evaluation using the HMF. 

\subsection{Predicting mass functions with the NN}
\label{sec:results_hmf}
It is now straightforward to compute the HMF from the U-Net output discussed above. First, we sum the class probabilities predicted by the NN over the entire box to obtain an (effective) number of particles $N_{p, i}$ per class $i$. Note that, in general, each particle distributes its mass over multiple bins according to the probabilities assigned by the NN. The approximate number of haloes in each class depends on the mass of a single Lagrangian fluid element ($M_p = M_\text{box} / N_p$, i.e.\ the total mass in the simulation volume divided by the number of fluid elements/particles). The estimated halo count for each bin can be obtained by calculating the total mass of all particles per class $i$ and dividing it by the average halo mass $\bar{M}_{\text{halo}, i}$: 
\begin{equation}
N_{\text{halo},i} = \frac{N_{p,i} M_p}{\bar{M}_{\text{halo}, i}} \;,
\end{equation}
where $\bar{M}_{\text{halo}, i}$ is the geometric centre of bin $i$. Class 0 can be ignored for this calculation as it represents the background. Finally, scaling by the box volume $V_{\text{box}}$, the HMF can be calculated as 
\begin{equation}
\text{HMF}_i = \frac{N_{\text{halo},i}}{V_{\text{box}} \Delta \ln{M}} = \frac{n_i}{\Delta \ln{M}} \;,
\end{equation}
where the halo number density $n_i = N_{\text{halo},i} / V_{\text{box}}$, and $\Delta \ln{M}$ is the logarithmic bin width. This allows us to use the CNN output directly and differentiably, without having to post-process the prediction. Furthermore, the use of probabilities naturally accommodates the inherent uncertainty of exact particle assignment in halo finders.

\begin{figure}
    \centering
    \includegraphics[width=\linewidth]{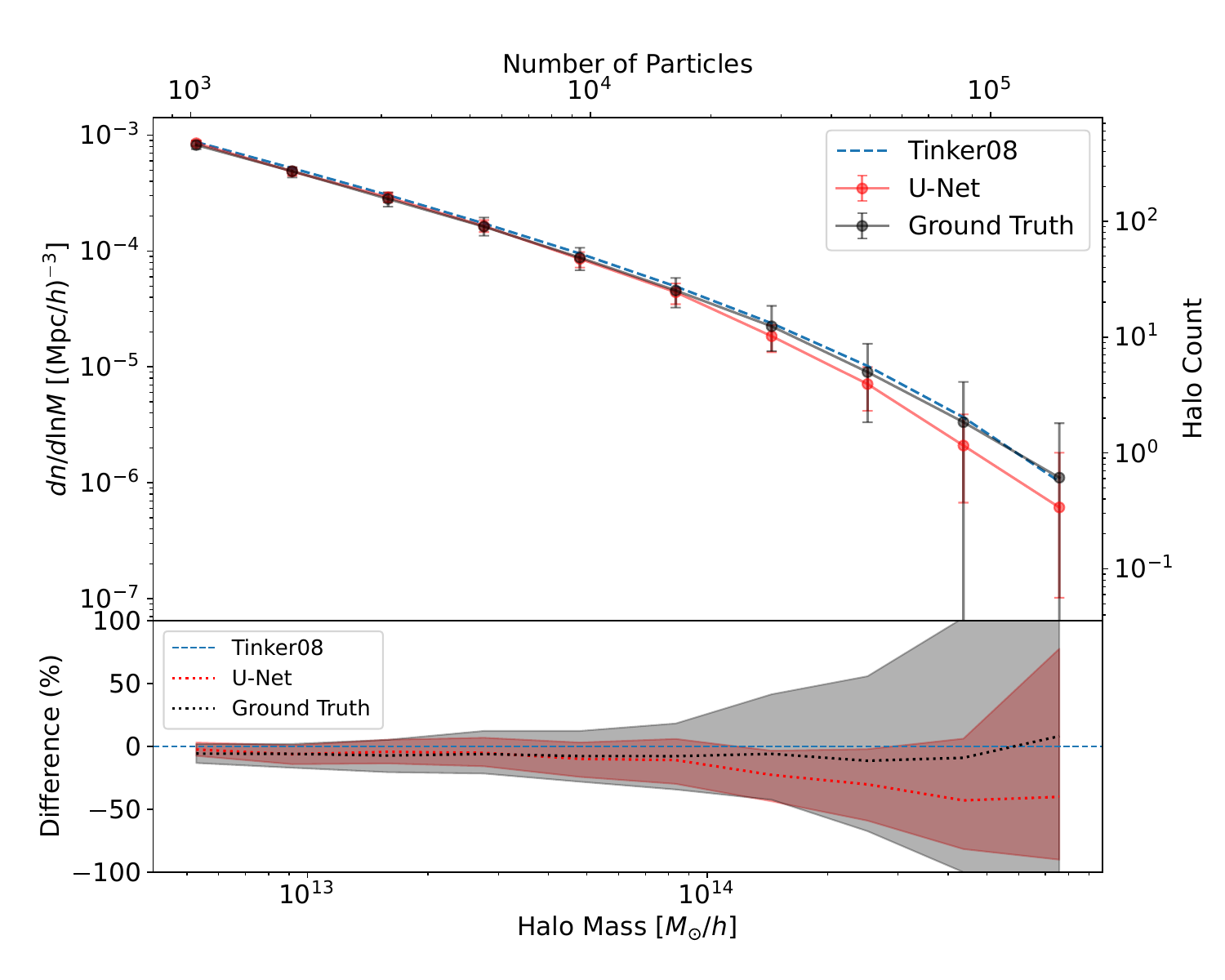} 
    \caption{The HMF for a fixed cosmology over 100 different realisations. The blue line shows the analytical Tinker08 HMF for the same cosmological parameters. Grey represents the ground truth and red the prediction. The dots in the upper plot show the mean value across all seeds with the error bars indicating the 5th to 95th percentile. The lower plot shows how far the true and predicted realisations deviate from the analytical expectation. The mean is indicated by the dashed line, while the shaded region shows the quantile range. The NN performs well for small and medium-sized halo masses, but somewhat underestimates higher mass haloes compared to the ground truth, which matches the Tinker08 line very well on average.}
    \label{fig:HMF_double-plot}
\end{figure}
In \autoref{fig:HMF_double-plot}, we compare the HMF for 100 different simulations and predictions (at fixed cosmological parameters) to reduce the effect of cosmic variance and analyse the averaged model output. Our fiducial cosmology for the following results is chosen as $\Omega_c = 0.260$, $n_s=0.968$, and $\sigma_8 = 0.809$. As we increase the number of test simulations, we expect the ground truth to align with the analytical Tinker08 prediction. The scatter over realisations increases as a function of halo mass -- as expected in view of the scarcity of cluster-size haloes. Interestingly, the quantile range of the NN predictions is smaller than that of the test simulations. We found that this is partially due to the probabilistic mass assignment. While the analytical solution and the true mean still fall within the error bars of the prediction, we can see that our NN, on average, underestimates the larger mass bins, for which there are $\lesssim 10$ haloes in our simulation volume.

Overall, our U-Net is able to predict the mass function at the $\approx 8$ \% level, though the reliability at the high-mass-end could be further improved as it is somewhat affected by cosmic variance due to the small box size employed.

\begin{figure*}
    \centering
    \includegraphics[width=\linewidth]{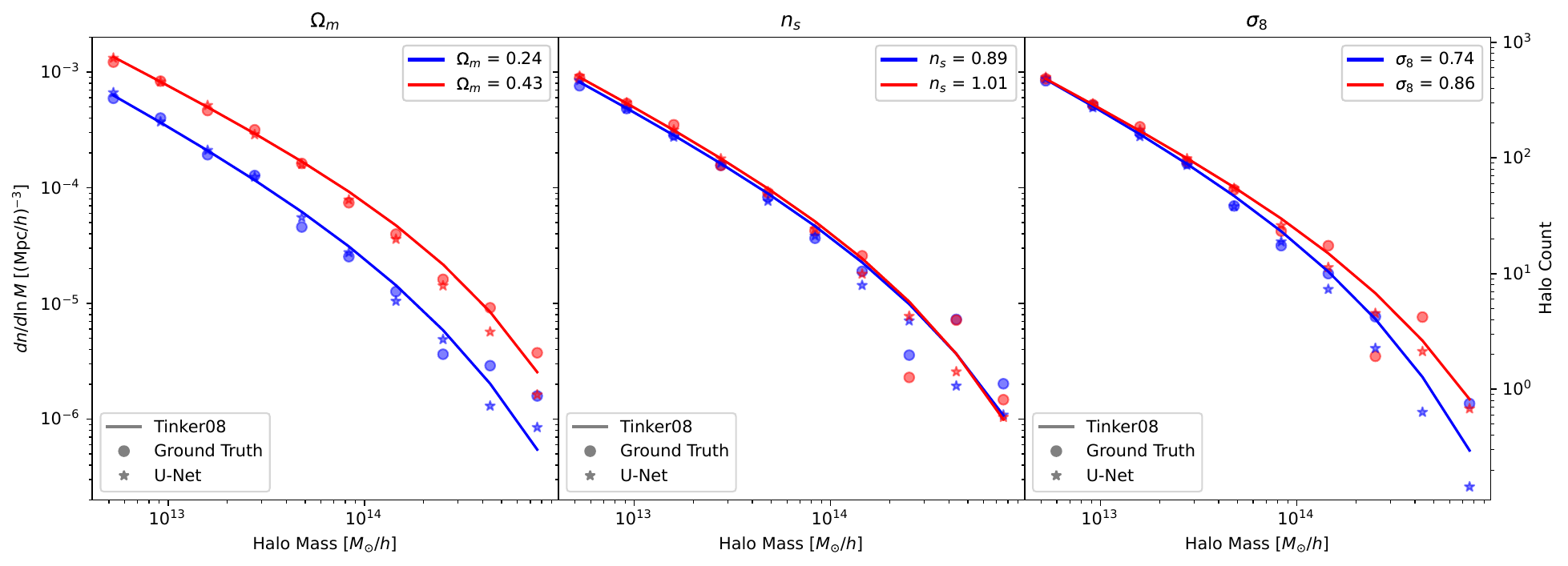}
    \caption{The HMF for the 20th (blue) and 80th (red) percentiles of each parameter value in the training Latin hypercube, while keeping all other parameters fixed. The lines show the expected average values based on the Tinker08 analytical HMF. The dots are the ground truth from a test simulation using $N$-body and \textsc{Rockstar}, with the model's mass bins. The stars represent the direct approach using the NN. Overall, the model is able to reproduce the analytical line and, more importantly, the ground truth quite well.}
    \label{fig:param-20_80}
\end{figure*}
\autoref{fig:param-20_80} illustrates how the NN-based HMF prediction responds to variations in each cosmological parameter, with all others (and the simulation seed) held fixed. The two values shown correspond to the 20th and 80th percentiles of the training range. Predictions based on the Tinker08 fit are shown as lines, the simulation ground truth as dots, and the NN predictions as stars. In the ideal case, the stars should align with the dots, but not necessarily exactly with the lines, as cosmic variance affects the number density to some extent, especially for higher-mass objects. We can see that the lower-mass end is recovered very accurately by the NN for all parameters. However, it deviates more towards larger haloes. Small differences in the counts are more noticeable there, since the absolute abundance of these objects is much lower. In the cases where the true simulation differs strongly from the analytical HMF, the network seems to slightly soften the effect. This is likely due to our probabilistic way of computing the halo counts. Nonetheless, even in that regime, the general trend is still captured quite well. Unlike $\Omega_m$, which has a positive effect on the HMF (i.e.\ a higher parameter value leads to an increase in the number density across all mass bins), increasing $n_s$ slightly reduces the abundance of high-mass haloes. This trend, with the lines crossing at $\approx 4 \times 10^{14} M_{\odot}/h$, is also predicted by our model. Meanwhile, a higher value of $\sigma_8$ increases the number density of larger haloes more than those with smaller masses. A higher amplitude of the primordial fluctuations leads to more haloes forming earlier and growing larger, which affects the high-mass end more \citep{Cui.2024.hmf-parameter-dependence-s8}. This dependence can also be seen quite clearly by our model prediction. The HMF being most affected by $\Omega_m$ and $\sigma_8$ also agrees with the literature (e.g.\ \citealt{Hung.2021.cluster-mass-function_parameters}).

\newpage
\section{Derivatives of mass functions w.r.t.\ cosmology and initial conditions}

\subsection{Validating cosmology dependence of the learned HMF}
\label{sec:results_autodiff}
We can now take our analysis a step further and compute the dependence of the mass function on cosmological parameters by computing the gradients of the NN output. Since our U-Net-based halo prediction pipeline is written in \textsc{Jax} \citep{jax.2018.github} and automatically differentiable by construction, gradients can be efficiently computed. While reverse-mode automatic differentiation (also known as backpropagation) is used during training (w.r.t.\ the trainable NN weights), we now employ forward-mode differentiation via \texttt{jax.jacfwd()} to compute derivatives w.r.t.\ cosmological parameters. While our \textsc{Jax}-based \textsc{Disco-Dj} $N$-body simulator is also automatically differentiable, this is not required here, as the HMF is predicted by the U-Net directly based on the linear density field.

The reference against which we compare these results is the simulation-based `ground truth', as well as the parametric Tinker08 HMF as implemented in the \textsc{Colossus} \textsc{Python} package \citep{Diemer.2018.COLOSSUS-HMF}, and the \textsc{Mira-Titan} emulator \citep{Bocquet.2020.mira-titan_HMF}. \textsc{Mira-Titan} is a non-parametric Gaussian process model based on $N$-body simulations. However, HMFs can be predicted only in terms of $M_{200c}$ instead of $M_{\text{vir}}$. We therefore perform a mass conversion with \textsc{Colossus}. Furthermore, we note that \textsc{Mira-Titan} is not able to emulate masses below $10^{13}M_{\odot}/h$. Since the \textsc{Mira-Titan} emulator is based on a Gaussian process, it might be possible to obtain derivatives via automatic differentiation -- something we do not pursue here, however. 

For both reference models and the `ground truth' values, we obtain gradients with a centred finite difference, given by
\begin{equation}
    \left( \frac{\partial f}{\partial \theta} \right)_{\text{finite}} = \frac{f(\theta + \epsilon) - f(\theta - \epsilon)}{2\epsilon} \;,
\end{equation} 
where $f$ is the HMF, $\theta$ represents a given cosmological parameter, and $\epsilon$ is chosen as $1\%$ of $\theta$. In the case of the ground truth this implies running two simulations for each partial derivative. The other parameters and the seed are left unchanged. We then compare the resulting partial derivatives to the automatic derivatives of the U-Net predictions.

\begin{figure*}
    \centering
    \includegraphics[width=\linewidth]{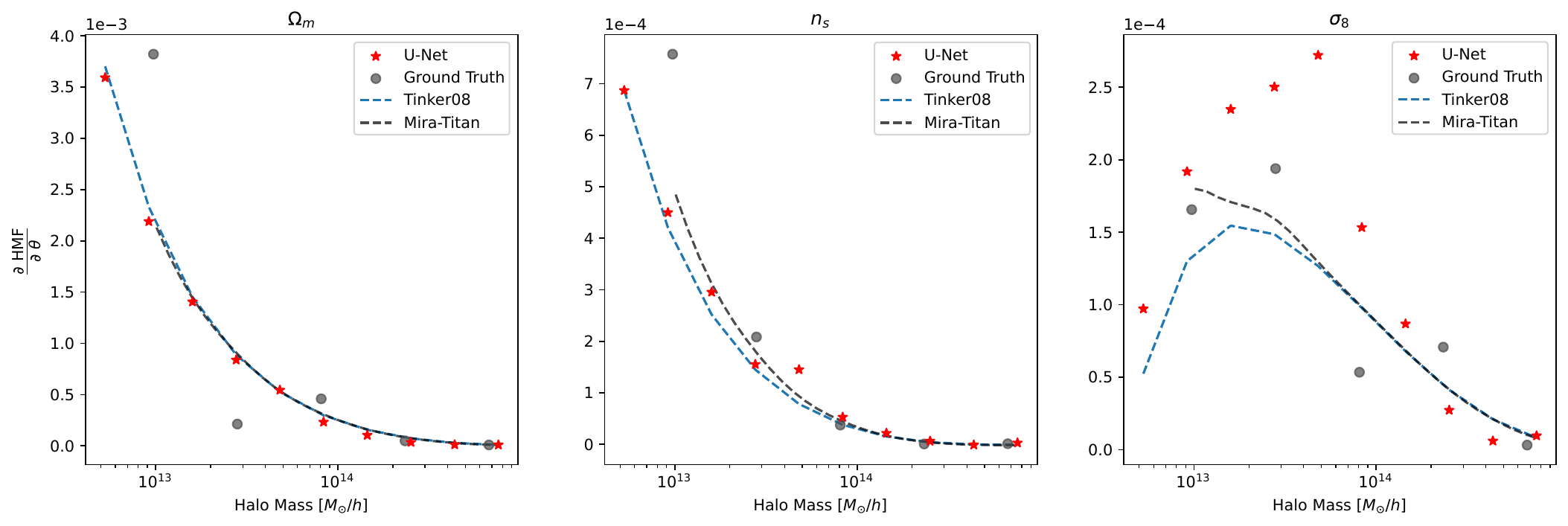}
    \caption{Derivatives of the HMF w.r.t.\ cosmological parameters for a single realisation. Tinker08 (blue dashed) and \textsc{Mira-Titan} (black dashed), as well as the ground truth (grey dotted) are based on a finite difference with $\epsilon=1\%$ of the parameter value. The NN prediction (red stars) is calculated via automatic differentiation from a single initial density field at the fiducial cosmology using the same fixed seed as the ground truth. We use fewer mass bins to decrease effects from cosmological variance in this individual seed. $\Omega_m$ (left) matches quite well, while $n_s$ (middle) shows a deviation between the analytical and the emulator HMF. For $\sigma_8$ (right), our U-Net predicts a somewhat larger effect than the ground truth for this realisation. }
    \label{fig:dHMF-dParams_truth}
\end{figure*}

In \autoref{fig:dHMF-dParams_truth}, we show the gradient of the HMF across all mass bins. Mira-Titan (solid black line) and Tinker08 (dashed black line) mostly agree on the parameter dependence, with slight differences at low masses for $n_s$ and $\sigma_8$. The simulation-based derivative (ground truth, black dots) for a single seed is very sensitive to cosmic variance. This leads to large scatter around the analytical prediction. In order to reduce this effect, we use fewer but wider mass bins in this plot. The resulting points still differ slightly from the lines, but the trend is clearly similar. Lastly, the derivatives from the U-Net (red stars) come directly from a single prediction using automatic differentiation. As seen above for the HMF itself, the probabilistic approach also dampens the effect of cosmic variance on its gradient. The result is that our model matches both the Tinker08 and the emulator well. The gradient w.r.t.\ $\Omega_m$ is captured almost perfectly. $n_s$ shows a small discrepancy between \textsc{Mira-Titan} and Tinker08, with our U-Net derivatives agreeing slightly better with the latter. However, the difference is quite small and we are only looking at a single prediction, so it is difficult to say this with certainty. Lastly, the derivatives obtained for $\sigma_8$ deviate more strongly. They overestimate the dependence for lower mass bins, but still capture a similar trend predicted by Tinker08. Interestingly, this parameter is also the one that shows the largest difference between the Tinker08 fit and the \textsc{Mira-Titan} emulator approach. 

\begin{figure*}
    \centering
    \includegraphics[width=\linewidth]{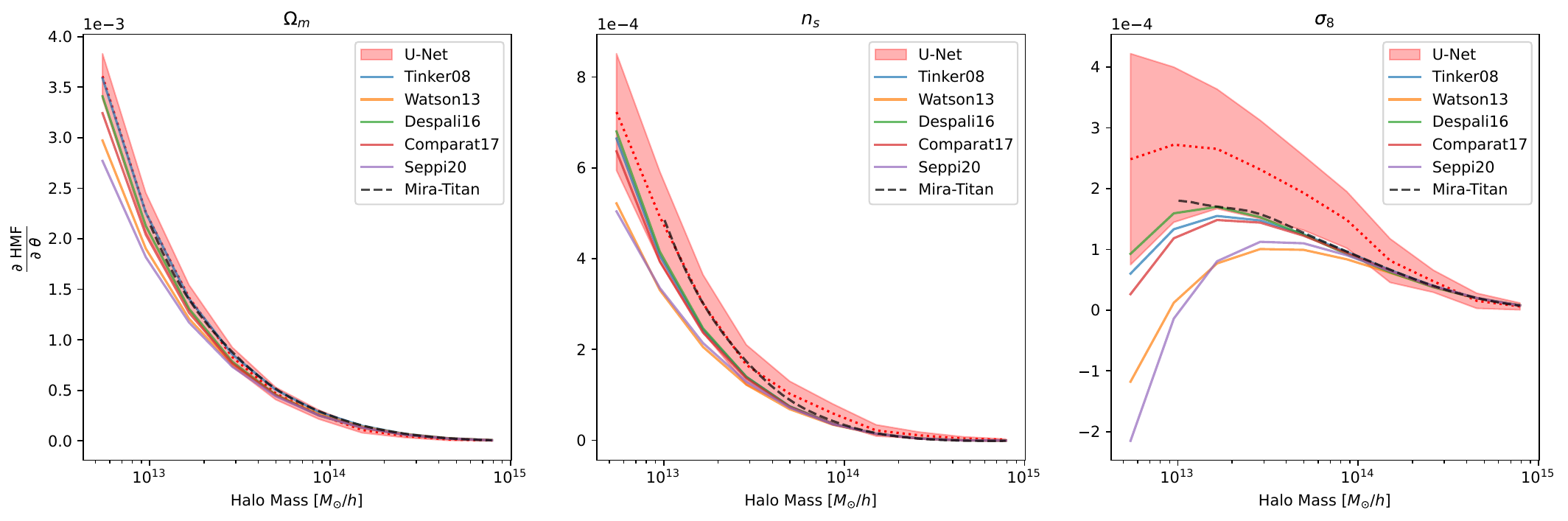}
    \caption{The dependence of the HMF on $\Omega_m$ (left), $n_s$ (middle) and $\sigma_8$ (right), comparing our NN (red) to a variety of analytical models (solid lines) and the \textsc{Mira-Titan} emulator (black dashed). The derivatives of the U-Net are calculated over 50 different seeds to reduce cosmic variance. The resulting mean is shown as the red dashed line with the shaded red region representing the $1 \sigma$ range. The difference between different analytical models and the \textsc{Mira-Titan} emulator (dashed black) is quite large, especially towards lower masses. The $1\sigma$ region of the U-Net fits quite nicely with a majority of these models, though it diverges more strongly for $\sigma_8$ at the low-mass end. However, this region also shows the largest discrepancies between the different models. 
    }
    \label{fig:dHMF-dParams_analyticals}
\end{figure*}

To further investigate this, we additionally evaluate the HMF derivatives by our U-Net across 50 independent realisations of the initial density field. In \autoref{fig:dHMF-dParams_analyticals}, we compare these gradients to \textsc{Mira-Titan}, as well as a range of analytical models from the \textsc{Colossus} package. While different parametrisations of HMFs are usually very similar, analysing the derivatives in this way demonstrates the variations in their sensitivity to different cosmological parameters. Surprisingly, there is quite a large disagreement between the analytical mass functions, especially at lower masses. They all converge to zero for more massive haloes, due to the absolute response becoming smaller. The discrepancy between the models is troubling, as it shows the underlying physical processes are not captured consistently. Since the \textsc{Mira-Titan} HMF emulator itself does not use any fitting function, but instead is directly based on a simulation suite to predict a mass function for a given cosmology, it can be argued, that it is more accurate. Out of the analytical functions, \citet{Watson:2013} and \citet{Seppi.2021.20hmf} diverge most notably from the rest -- especially at low masses for $\sigma_8$, where they are the only to predict a negative derivative. At the same time, Tinker08 and \citet{Despali:2016} seem to match \textsc{Mira-Titan} best. 

Our average NN gradient seems to agree extremely well with the emulator for $\Omega_m$ and $n_s$, showing better agreement than the analytical HMFs. As for $\sigma_8$, it overestimates the effect of the parameter on low-mass haloes, though \textsc{Mira-Titan} still falls right on the edge of the $1 \sigma$ (shaded) region. Interestingly, $\sigma_8$ also shows the largest disagreement between the analytical predictions so it might simply be more difficult to capture accurately. All in all, our NN produces competitive results and can be used in a differentiable manner reliably.

\subsection{Extrapolation by auto-differentiation of parametric mass-functions}
\label{sec:results_taylor}
In this section, we showcase a possible application of differentiable mass functions. Specifically, we propose to enhance arbitrary (potentially non-differentiable) HMFs from a base model with analytical gradients provided by an autodifferentiable model. While the gradients will only approximately align with those of the base model (which are not accessible), we will show below that, for small deviations from a fiducial cosmology, the accuracy of the approximation is excellent. This enables the application of gradient-based methods such as HMC -- for which, thanks to the Metropolis--Hastings step, the use of approximate gradients does not bias the sampling, although the sampling efficiency may suffer with inaccurate gradients. 
While we could utilise our NN pipeline for the derivatives, we instead opt to implement an analytical HMF in \textsc{Jax}, as they are independent of any particular realisation and generally better understood. 

In what follows, we briefly describe how we construct such an autodifferentiable analytical HMF. A key requirement towards this goal is the ability to autodifferentiably produce linear power spectra. While we could use the \citet{Eisenstein1998BaryonicFunction} parametrisation as above, we now build a differentiable HMF estimator onto the recently presented \textsc{Disco-EB} Einstein--Boltzmann solver \citep{Hahn:2024:discoeb}. The difference between the methods is within $\approx 6\%$ for the HMF in our mass range. 

The linear power spectra produced by \textsc{Disco-EB} are evaluated at scale factor $a=1$ and normalised to match a specified $\sigma_8$, before the mass variance $\sigma(M)$ is calculated using a top-hat filter in Fourier space. We implemented the \citet{Tinker.2008.hmf} HMF as our fitting function, though other analytical solutions are equally possible. Its multiplicity function is parametrised as 
\begin{equation} \label{eq:tinker_f}
    f(\sigma) = A \left[ \left(\frac{\sigma}{b} \right) ^{-a}  +1 \right] e^{-c/\sigma^2} \;,
\end{equation}
with
\begin{equation} \label{eq:tinker_sigma}
    \sigma = \int P(k) \hat{W}(kR)k^2 \, dk \;,
\end{equation}
where $\hat{W}$ is the Fourier transform of the top-hat filter of radius $R$. $A$, $a$, $b$, and $c$ are constants that were fit to match simulation data.

Since Tinker08 is based on $\delta_{200m}$, we interpolate the coefficients to $\delta_{\text{vir}}$. The final HMF is fully differentiable w.r.t.\ cosmological parameters. Note that while in the previous sections we restricted ourselves to varying $\Omega_m$, $n_s$, and $\sigma_8$, we now additionally allow for time-varying dark energy parametrised by the Chevallier--Polarski--Linder model $w(a) = w_0 + (1 - a) w_a$ \citep{chevallier2001accelerating, PhysRevLett.90.091301} and variable neutrino mass $m_\nu$.

To validate the precision of our prediction of the mass function itself, we compared our implementation of Tinker08 using \textsc{Disco-EB} against \textsc{Colossus} and found a deviation of $< 2\%$, which we deemed sufficiently small, considering that using a different model or changing the cosmology causes a much more significant difference. 

\begin{figure}
    \centering
    \includegraphics[width=\linewidth]{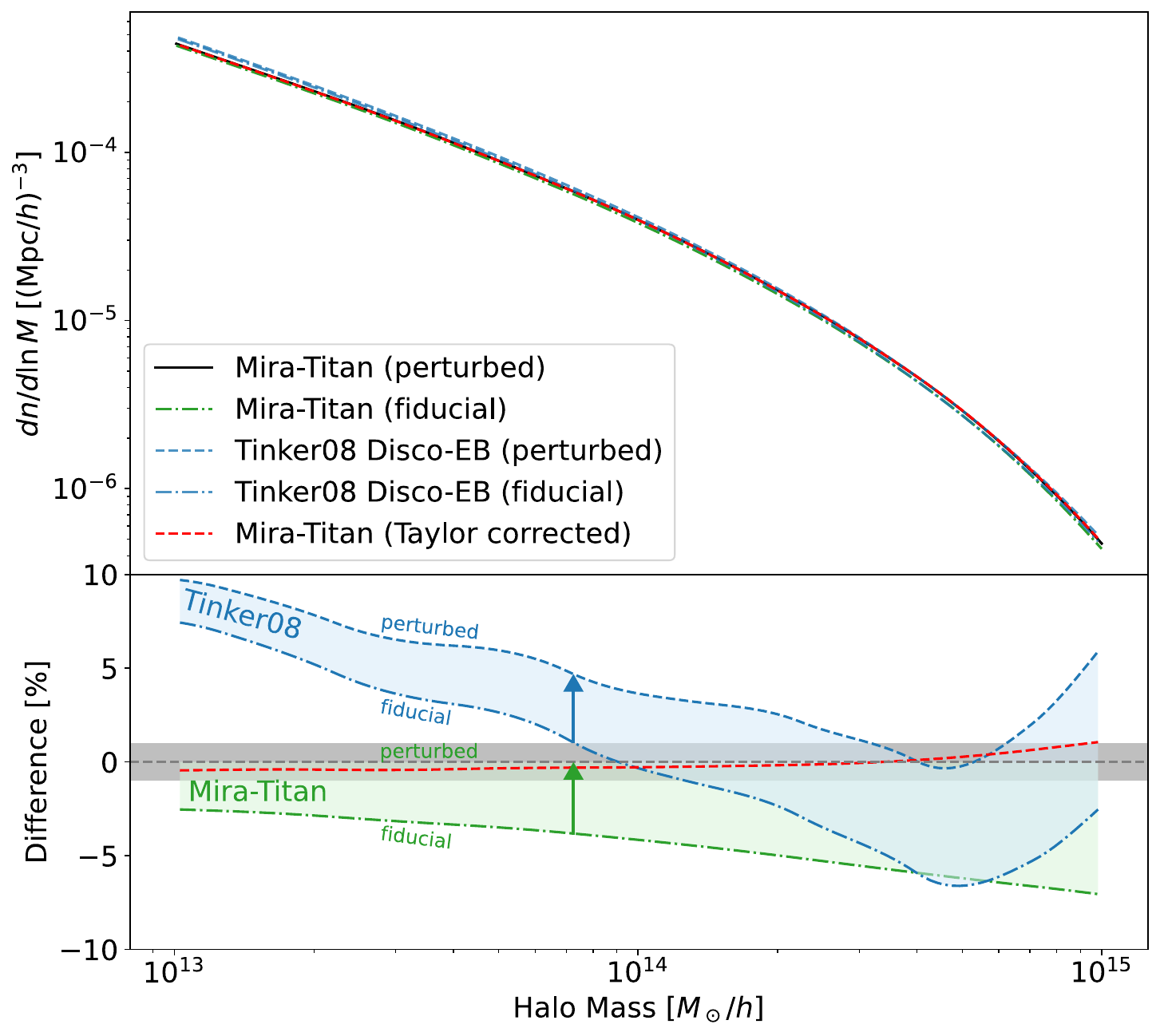}
    \caption{The HMF for two different cosmologies comparing \textsc{Mira-Titan} and Tinker08. The following parameters are increased by $1\%$ for the perturbed case: $\Omega_m$, $n_s$, $\sigma_8$, $w_0$, $w_a$, and $m_{\nu}$. The difference between fiducial cosmology (dash-dot) and perturbed (dashed) is highlighted with a shaded region for each model. The fiducial-cosmology \textsc{Mira-Titan} HMF is used as a baseline for the lower deviation plot. Blue lines show the analytical Tinker08 case, while black and green correspond to the \textsc{Mira-Titan} emulator. At low masses, the deviation between these models at fixed cosmology is larger than the shift induced by the cosmology itself. However, since the \textit{response} of both models to the cosmology shift is similar, Taylor correcting \textsc{Mira-Titan} using the derivatives from the Tinker08 HMF yields an accurate HMF prediction for the perturbed cosmology (red line).
    }
    \label{fig:MT-taylor}
\end{figure}
 
We choose the \textsc{Mira-Titan} emulator as our base model and augment it with derivatives from our autodifferentiable Tinker08 implementation. Specifically, we apply a simple first-order Taylor expansion around a fiducial cosmology using the gradients extracted from Tinker08 to approximate the \textsc{Mira-Titan} HMF of a perturbed cosmology like this:

\begin{equation} \label{eq:taylor_correction}
\begin{aligned}
    \mathrm{HMF}^\mathrm{MT}(\theta) &\approx \mathrm{HMF}^\mathrm{MT}(\theta_\mathrm{fid}) \\
    & \quad \left[ 1+ \left( \frac{\theta - \theta_\mathrm{fid}}{\theta_\mathrm{fid}} \right) \left. \frac{\partial \ln \mathrm{HMF}}{\partial \ln \theta} \right|_{\mathrm{fid}} ^{\mathrm{T08}} \right] \;,
\end{aligned}
\end{equation}
where MT and T08 represent \textsc{Mira-Titan} and Tinker08 respectively. 

\autoref{fig:MT-taylor} shows the quality of this linear approximation. The \textsc{Mira-Titan} HMF for the perturbed cosmology deviates from the baseline case by $> 5\%$ for large halo masses, and exactly this deviation is what we aim to capture with the Tinker08 gradients. Indeed, although the Tinker08 HMF for the perturbed cosmology itself is off by roughly $10\%$ for small halo masses, the linear Taylor approximation in \autoref{eq:taylor_correction} using Tinker08 derivatives w.r.t.\ all five considered parameters reproduces the perturbed \textsc{Mira-Titan} HMF at the sub-percent level. This demonstrates that the Tinker08 gradients are similar enough to \textsc{Mira-Titan}, as can be seen by the width of the shaded regions. This allows them to be used in cases where they would be necessary but unavailable for \textsc{Mira-Titan}, such as parameter inference.

\subsection{Dependence of HMFs on the white noise amplitude}
\label{sec:results_whitenoise}

\begin{figure}
    \centering
    \includegraphics[width=\linewidth]{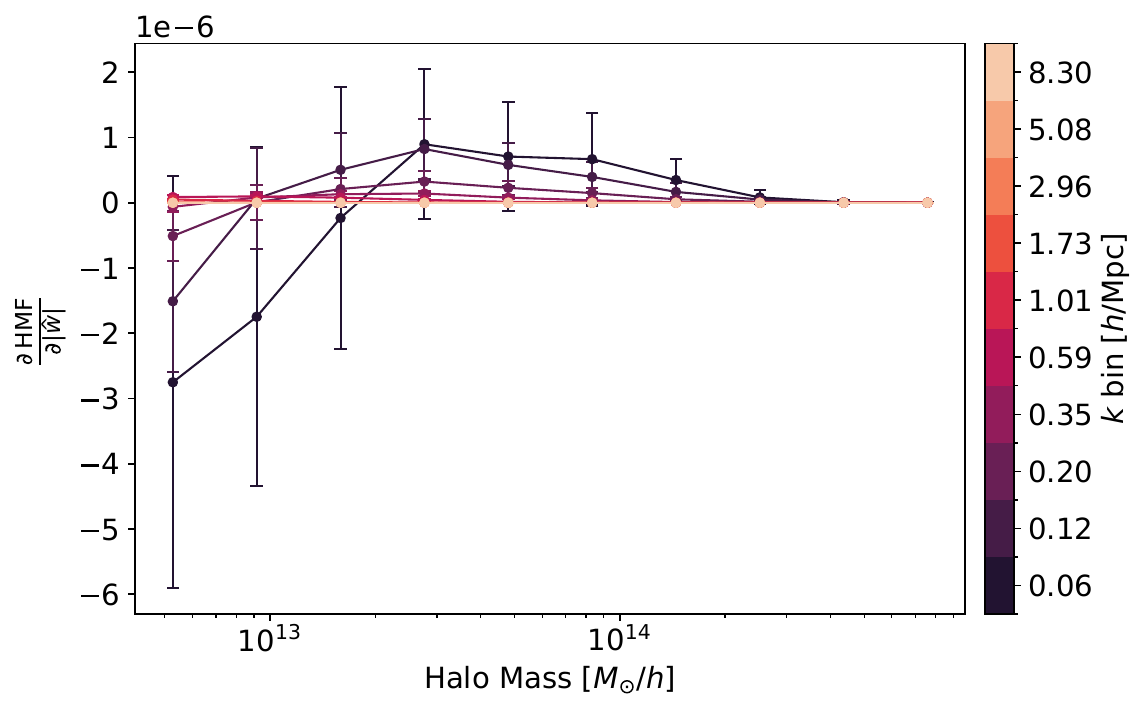}
    \caption{The derivative of the HMF w.r.t.\ the white noise field calculated using automatic differentiation for our NN, evaluated at $|\hat{w}| \equiv 1$. Each line represents a radially binned $k$ mode for every mass bin of the U-Net predictions. The derivatives are calculated over $50$ independent realisations, each with randomly drawn phases. The error bars indicate the standard deviation. 
    }
    \label{fig:dHMF-dw}
\end{figure}
Having studied the sensitivity of the HMF w.r.t.\  cosmological parameters, we now turn to its dependence on the initial density fluctuations. Specifically, we consider a linear density field in Fourier space
\begin{equation}
    \hat{\delta}_{\mathrm{lin}}(\boldsymbol{k}) = \hat{w}(\boldsymbol{k}) \sqrt{P_{\mathrm{lin}}(k)} \;,
\end{equation}
where $w \in \mathcal{N}(0, 1)$ is a white noise realisation, $P_{\mathrm{lin}}(k)$ is the linear power spectrum, and carets indicate Fourier-transformed quantities. Our goal is to analyse how infinitesimal changes in the white noise amplitude $|\hat{w}(\boldsymbol{k})|$ affect the resulting HMF prediction by our U-Net. In the discrete setting, $\hat{w}(\boldsymbol{k})$ is represented by $N^3 = 256^3$ complex values corresponding to discrete wave vectors. For such high-dimensional inputs, reverse-mode automatic differentiation, as provided by the \texttt{jax.jacrev()} method, is ideally suited for computing derivatives, as the Jacobian of the HMF w.r.t.\ the white noise field is extremely wide and short -- with $n_\mathrm{bin}$ rows and $N^3$ columns.
With the reverse mode, the Jacobian is constructed efficiently row-by-row (in contrast to the forward mode, which builds it column-by-column and is better suited for low-dimensional inputs, such as cosmological parameters). Since we expect the derivative w.r.t.\ the white noise \textit{modulus} to be the most informative, we compute gradients only w.r.t.\ $|\hat{w}|$, evaluated at its expected value of unity for each mode, assigning each mode a random phase uniformly drawn in $[0, 2 \pi)$.

After propagating gradients through our U-Net back to the white noise field, we radially bin the resulting derivatives in Fourier space into $k$-bins. 
The binned sensitivities are shown in \autoref{fig:dHMF-dw}, where each curve corresponds to a different $k$-bin, and the $x$-axis represents halo mass. Darker colours show derivatives w.r.t.\ larger-scale modes (i.e.\ smaller $k$). The plot reveals that large-scale fluctuations have the most significant impact on the HMF. Increasing the white noise amplitude on large scales reduces the abundance of low-mass haloes, while enhancing the formation of intermediate-mass haloes. This behaviour is consistent with the intuitive picture that large-scale overdensities promote the hierarchical merging of small haloes into larger structures.
In contrast, smaller-scale modes (red curves) yield strictly positive gradients across the entire mass range but with much smaller magnitudes. Our result agrees with earlier studies, such as \citet{Little.1991.scale-sensitivity-mass}, who similarly found that large-scale modes affect the structure formation to a far more significant degree than small scales. This experiment demonstrates that our model not only captures the sensitivity of the HMF to global cosmological parameters, but also supports differentiability w.r.t.\ the specific realisation of the initial conditions. This opens up the possibility of studying field-level responses and enables the use of inference strategies where initial conditions are treated as latent variables.

\section{Conclusion}
In this paper, we train a multiclass semantic classification network to predict the mass of protohalo patches from the linear density field. We use this as an automatically differentiable way to obtain halo catalogues, with the main focus on the resulting HMF, along with the dependence on cosmology parameters or properties of the initial random field. We generate a large training set of fast cosmological $N$-body simulations based on the GPU-based \textsc{Disco-Dj} code, and use \textsc{Rockstar} to identify haloes. We then trace their particles back to their Lagrangian positions and label them according to their final halo mass, which is the precise task that we want a neural network to perform: to label Lagrangian protohalo patches by their final halo mass. Specifically, we use a U-Net architecture to predict 12 classes, with the inner 10 corresponding to logarithmically spaced masses between $4 \times 10^{12} M_{\odot}/h$ and $1 \times 10^{15} M_{\odot}/h$, and the outer two containing masses above and below this range. The fully trained NN is rather conservative (i.e.\ it reaches a higher TNR than TPR), but performs overall well with an F$_1$-score of $\approx77\%$. The resulting HMF is accurate at an average of $\approx 8\%$. Lower mass haloes are predicted more reliably, while the U-Net underestimates the high-mass end due to the extremely low halo counts in our training data. This latter aspect could be rectified by training on larger volumes. We then use automatic differentiation of the trained model to directly calculate the dependence of the HMF on three different cosmological parameters in order to validate how well the network has captured cosmological information content. In comparison to finite difference derivatives from both simulations and literature models, the derivatives for $\Omega_m$ and $n_s$ are captured extremely well, but deviate more strongly for $\sigma_8$. We also show how the derivatives of the HMF can be used as corrections for non-differentiable HMFs such as \textsc{Mira-Titan} \citep{Bocquet.2020.mira-titan_HMF}, achieving an accuracy at the sub-percent level. 
We achieve this by performing an extrapolation of a reference prediction at a fiducial cosmology (in our case \textsc{Mira-Titan}) based on the gradients of another model (in our case, we used the Tinker08 parametric model, but the U-Net could also be used). 
The result demonstrates that automatically differentiable HMFs can be used as gradient information for parameter inference even if the baseline HMF model is not. What is more, this approach allows investigating e.g.\ the response of \textsc{Mira-Titan} to changes in cosmological parameters that are not part of the parameter set for which the emulator has been trained.

Lastly, since our U-Net produces an HMF for a given linear density field realisation, we can also compute the response of the HMF to variations in the white noise Gaussian random field, which could be used e.g.\ for estimating uncertainties due to cosmic variance.
We find (as expected) that more power in the initial conditions on large scales increases the abundance of medium-sized haloes while decreasing the number of small haloes.

The results presented here represent a first step towards a fully data-driven approach for assigning halo (and ultimately galaxy) properties to protohalo patches and constructing summary statistics that capture both the mutual dependence of these properties and their interdependence with the underlying cosmological parameters. Our results indicate that our proof-of-concept NN-based model is accurate at the few-percent level and captures cosmological dependence reasonably well. Natural next steps to further improve upon our model could include (1) moving beyond binned labels to instead predict continuous halo properties, (2) a `windowed labelling' approach allowing to label much larger volumes, (3) more rigorous tests of non-universality of the predicted mass function (for example, we currently do not provide growth factors and growth rates as inputs for the NN), (4) predict mass functions for multiple mass definitions, and (5) extend our framework toward an entirely differentiable halo identification.

\section*{Acknowledgements}
Calculations discussed in this article were performed using supercomputer resources provided by the Vienna Scientific Cluster (VSC). We thank Thomas Flöss for help with the Fourier up- and down-sampling, as well as the idea for implementing multi-GPU training. Furthermore, we thank Jens Stücker for helpful discussions and Fiona Sonnek for the visualisation of the U-Net architecture in \autoref{fig:cnn_architecture}. 

\bibliography{sources}

\vspace*{0.2cm}

\clearpage
\appendix
\begin{figure*}
    \centering
    \includegraphics[width=\linewidth]{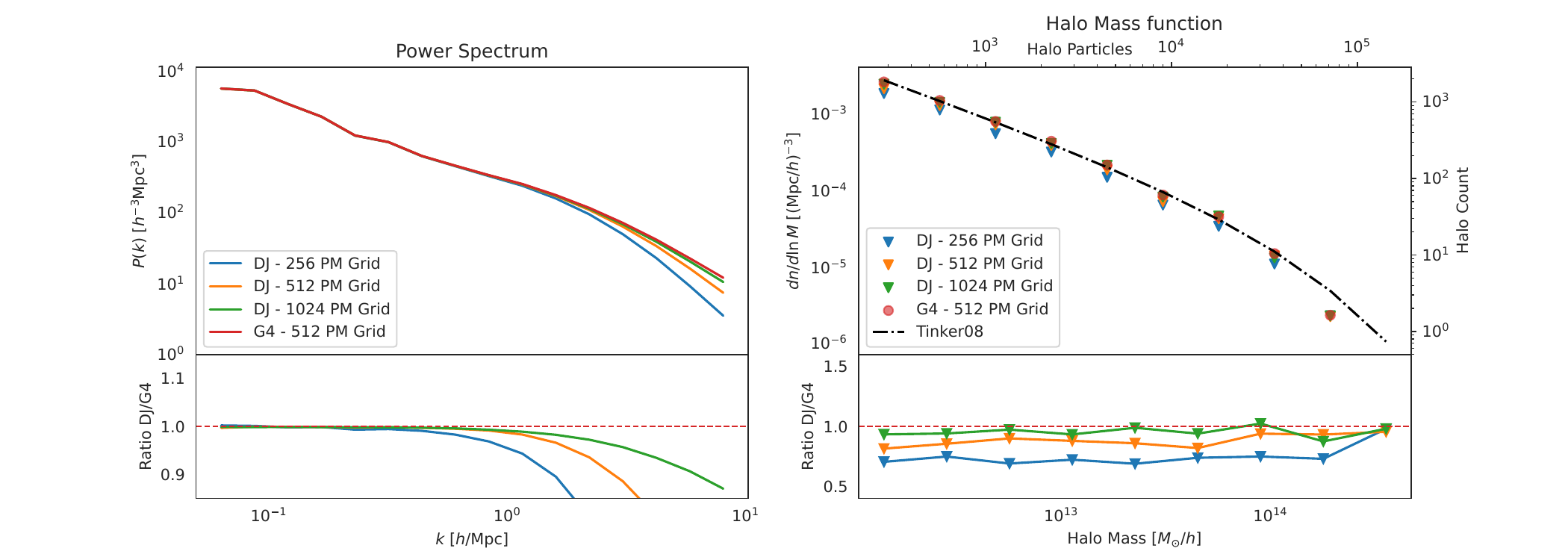}
    \caption{The effect of varying the particle mesh (PM) resolution in \textsc{Disco-Dj}. We run a \textsc{Gadget-4} simulation as a baseline using the same initial conditions and particle number ($256^3$). The left plot shows the power spectrum at redshift $z = 0$, with the bottom representing the ratio compared to \textsc{Gadget-4}. Increasing the grid resolution improves the power on small scales. The right plot shows the resulting HMF for these four different simulations, as well as the analytical Tinker08 line. The bottom ratio is again relative to the \textsc{Gadget-4} run, since any specific seed will deviate from the analytical prediction due to cosmic variance. Using four times the particle resolution per linear dimension for the PM grid in \textsc{Disco-Dj} yields the best results again.}
    \label{fig:dj_vs_g4}
\end{figure*}

\section{A. Convergence of N-body simulation results}
\label{sec:app:convergence_nbody}
In \autoref{fig:dj_vs_g4}, we analyse the effect of the PM resolution on the power spectrum and HMF while keeping the number of particles (and initial seed) fixed. We compare our \textsc{Disco-Dj} results to a standard \textsc{Gadget-4} simulation. The latter uses TreePM, which splits the forces into a long-range part (solved with the PM method) and a short-range part (solved with a tree) and therefore achieves higher force resolution than a pure PM code. These short-range forces are important for halo formation, and underestimating them causes the haloes to be generally smaller, which shifts the HMF to lower number densities. In order to improve the force calculations at these scales, we can increase the PM grid resolution. The left panel shows that increasing the number of grid cells shifts the drop in power relative to \textsc{Gadget-4} to smaller scales. We also checked that the cross-spectrum between \textsc{Gadget-4} and \textsc{Disco-Dj} (1024 PM) is 98.8\% at the Nyquist scale $k_{\mathrm{Nyq}}$. Similarly, a larger PM resolution leads to a more accurate HMF (right plot) in our setting. However, a higher grid resolution at fixed particle number does not always increase the power spectrum; for larger simulation boxes, enhanced discreteness due to the small particle number per cell can lead to a decrease instead.

Since we found a high PM resolution to improve the results in our case, we used a 1024 PM grid for all our simulations in this paper. While it is to be assumed that further increasing the resolution would improve the short-range forces even more, we reach a memory limitation on the GPU. Fortunately, four times the particle number per dimension seems to be sufficient for our purpose. 

\begin{figure}
    \centering
    \includegraphics[width=0.5\linewidth]{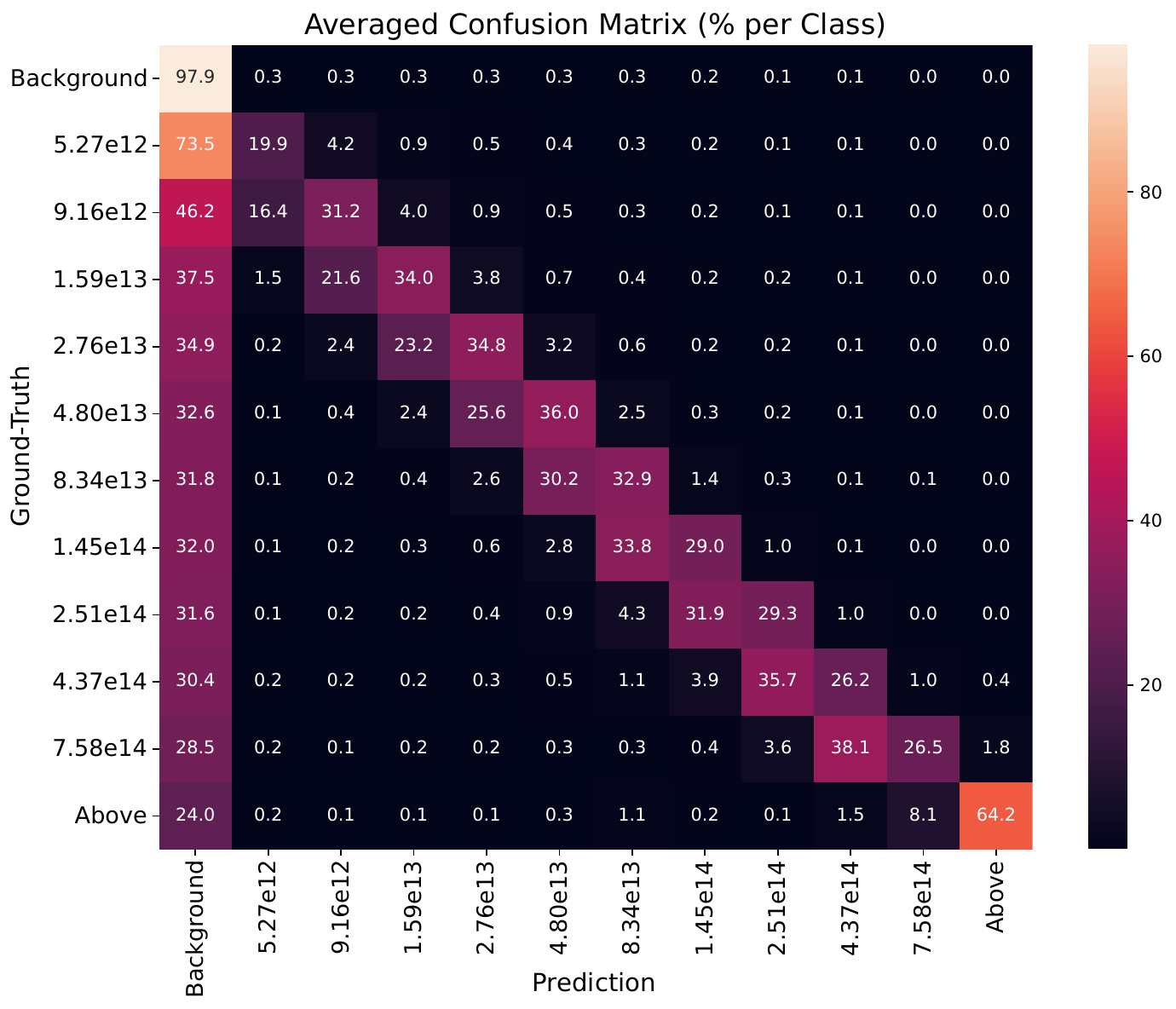}
    \caption{Averaged results from 100 test simulations showing how often each mass class is predicted correctly. The predictions correspond to the class with the highest probability (i.e.\ the $\operatorname{argmax}$ over classes). Class labels indicate the mean mass in $\msun /h$. Exceptions are the lowest bin, which represents the background, and the highest bin, which contains all haloes above $1 \times 10^{15} M_{\odot}/h$. A perfect model would have 100\% along the diagonal and zeroes elsewhere. True background voxels are accurately identified, and the probability of mistaking a true halo voxel for background generally decreases with increasing halo mass. Most halo voxels are correctly classified or assigned to the adjacent lower-mass bin.  
    }
    \label{fig:confusion_matrix}
\end{figure}

\section{B. Prediction accuracy of the U-Net}
\label{sec:app:confusion}
In \autoref{fig:confusion_matrix}, we show how accurate our NN is across each individual class. As seen in our previous tests, our U-Net identifies background voxels with a very high accuracy of 97.9\%. The vast majority of NN predictions for true halo voxels fall into one of three categories: 1) correct classification, 2) false negatives (i.e.\ predicted as background), which are to some extent expected near halo boundaries, and 3) one bin too low (as indicated by the rather high values on the lower off-diagonal of the confusion matrix). Regarding the last category, since the difference between adjacent classes is small, such slight misclassification is not particularly concerning. Notably, \autoref{fig:confusion_matrix} shows only the most likely predicted class (i.e.\ the class with the highest predicted probability, given by the $\operatorname{argmax}$). Therefore, even when the predicted class is off by one bin, the NN may still assign a substantial probability to the correct class. Importantly, it is extremely rare that our CNN predicts a mass that is further off the diagonal for a protohalo patch. This shows that incorporating a physical meaning to the distance between classes through the use of the EMD loss worked as intended.

\section{C. Toy example: Why the EMD$^2$ loss avoids spurious interpolation through low-mass bins}
\label{sec:app:emd-toy}
\begin{figure*}
    \centering
    \includegraphics[width=\linewidth]{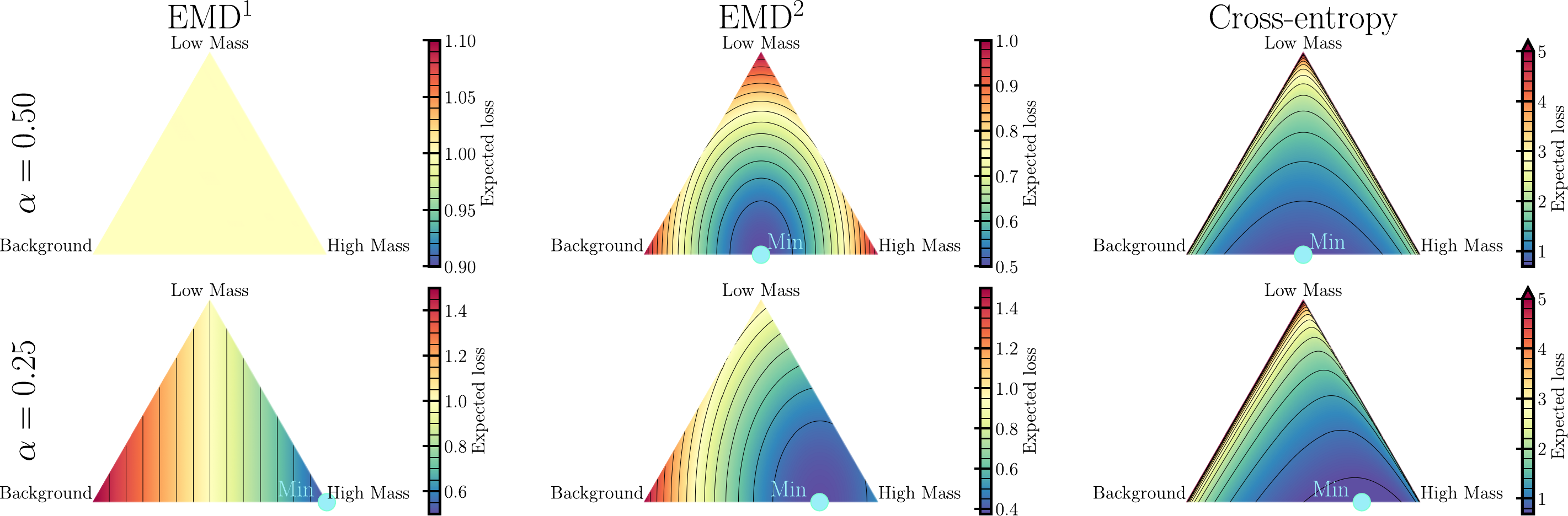}
    \caption{Expected loss over the 3-class prediction simplex for a toy classification problem with classes corresponding to background (bin~0), low-mass halo (bin~1), and high-mass halo (bin~2). The ground truth class is either background (with probability $\alpha$) or high-mass (with probability $1 - \alpha$). The upper and lower row correspond to $\alpha = 0.5$ and $\alpha = 0.25$, respectively. The left and central panels show the expected squared Earth Mover’s Distance loss with exponent $1$ and $2$, and the right panels show the standard (bin-wise) cross-entropy loss. The EMD$^1$ loss is constant over the simplex for $\alpha = 0.5$ and is minimised by $\bm{p} = (0, 0, 1)^\top$ for $\alpha < 0.5$ (and $\bm{p} = (1, 0, 0)^\top$ for $\alpha > 0.5$). For EMD$^2$ and cross-entropy, the expected loss is minimised at $\bm{p} = (\alpha, 0, 1 - \alpha)^\top$, which places probability on the two plausible outcomes in proportion to their occurrence frequency and assigns no weight to the unsupported intermediate class. This explains why our NN does not spuriously assign low-mass halo bins to halo edges, but rather maintains a consistent prediction all the way out to the halo edge. The cross-entropy loss diverges at the edges opposite the true classes due to vanishing probabilities.}
    \label{fig:emd-toy}
\end{figure*}
In the main body, we found the NN predictions to be remarkably robust. A consistent mass is typically assigned to all pixels of a protohalo patch, even near the halo edge, see \autoref{fig:multiple_predictions}. Since we use the EMD$^2$ loss with a one-dimensional ground-distance embedding -- rather than a bin-wise loss function that does not assume any bin ordering such as the cross-entropy loss -- low-mass bins lie between the background and high-mass bins. Therefore, it might be surprising at first that the predictions of high-mass haloes do not interpolate across low-mass bins to the surrounding background when gradually moving outwards starting from the centre of a halo patch.

To better understand the behaviour of the EMD$^2$ loss in a situation where the NN is confident that a voxel lies either in a high-mass bin or in the background, we study a toy scenario involving three classes: 0 -- background, 1 -- low-mass halo, and 2 -- high-mass halo. In particular, we aim to understand why interpolating through the low-mass bin would not be beneficial for the NN in terms of the loss in such a scenario. We consider a single voxel and model the classifier output as a probability vector $\bm{p} = (p_0, p_1, p_2)^\top$, where $p_i$ denotes the predicted probability of class $i$, and $\sum_i p_i = 1$.

We assume that the ground truth class is either background ($i = 0$), with a true probability of $\alpha$, or high-mass halo ($i = 2$), with a true probability of $1 - \alpha$. The expected EMD$^2$ loss is then given by:
\begin{equation}
\mathbb{E}[L_{\mathrm{EMD}^2}] = \alpha \, \mathrm{EMD}^2(\bm{p}, \bm{e}_0) + (1 - \alpha) \, \mathrm{EMD}^2(\bm{p}, \bm{e}_2) \;,
\end{equation}
where $\bm{e}_i$ denotes the one-hot target vector for class $i$, and $\mathrm{EMD}^2(\bm{p}, \bm{e}_i)$ is the squared EMD between the prediction $\bm{p}$ and $\bm{e}_i$ as defined in \autoref{eq:emd}.

In \autoref{fig:emd-toy}, we evaluate this expected loss for $\alpha = 0.5$ and $\alpha = 0.25$ over the 2-simplex (i.e.\ all valid predictions $\bm{p}$ in $[0, 1]^3$ that sum up to unity) and visualise the result in barycentric coordinates, with each vertex corresponding to a 100\% confident prediction for one of the three classes. The colour of each point indicates the expected loss, and isolines are added to guide interpretation.

We first discuss the symmetric case of $\alpha = 0.5$. Here, the minimum expected EMD$^2$ loss occurs at $\bm{p} = (0.5, 0, 0.5)^\top$, corresponding to a prediction that assigns equal weight to the two possible true classes and no weight to the intermediate low-mass bin. In this case, one finds $\mathrm{EMD}^2(\bm{p}, \bm{e}_0) = \mathrm{EMD}^2(\bm{p}, \bm{e}_2) = 2 \times 0.5^2 + 0 = 0.5$, leading to an expected EMD$^2$ loss of $0.5$ when averaging over the two outcomes. 
Notably, placing all probability in the central bin, i.e.\ $\bm{p} = (0, 1, 0)^\top$, yields the \textit{maximum} expected EMD$^2$ loss of 1: the entire squared probability mass of 1 must always be transported by one bin (from bin 1 to the true bin 0 or 2), regardless of the outcome. This expected loss is the same as for the other two corner predictions, $\bm{p} = \bm{e}_0$ or $\bm{p} = \bm{e}_2$: in those cases, the prediction is correct half the time and two bins off the other half -- also yielding a loss of 1. Uniformly distributing the probability mass over the three classes, i.e.\ $\bm{p} = (1/3, 1/3, 1/3)^\top$ (the centre of the triangle), yields a smaller expected loss of $5/9$, which is however still greater than for the optimal choice $\bm{p} = (0.5, 0, 0.5)^\top$. This confirms that, despite the imposed ordinal structure, the EMD$^2$ loss does not encourage interpolation through intermediate bins unless they are actually supported by the data. Due to the symmetry of the scenario, the penalty for deviating from the optimal prediction is the same in the directions of the background and the high-mass bin.

Note that the exponent $q$ in the EMD loss is crucial here: for the EMD$^1$ case (i.e.\ $q = 1$), the loss is constant over the prediction simplex for $\alpha = 0.5$. Intuitively, this difference between EMD$^1$ and EMD$^2$ comes from the fact that for EMD$^1$, moving half the probability mass by twice the distance leaves the loss unaffected, while for the strictly convex EMD$^2$, the former case is beneficial (as the difference between the CDFs is squared and thus becomes smaller for values in $(0, 1)$ in \autoref{eq:emd}). 

For comparison, we additionally show the same setup using the standard cross-entropy loss, given by 
\begin{equation}
    L_X(\bm{p}, \bm{t}) = -\sum_{i=1}^C t_i \ln(p_i) \;,
\end{equation}
which in a realistic scenario would be averaged over the voxels similarly to \autoref{eq:emd}.
Also in this case, the loss is minimised for $\bm{p} = (0.5, 0, 0.5)^\top$ -- which is less surprising, as the bins are treated independently and any probability assigned to the low-mass bin is wasted given that $t_1 = 0$ for both possible outcomes. However, the geometry of the isolines differs from the EMD$^2$ case -- most notably, they are much flatter near the edges opposite the background and high-mass corners. This is because approaching these edges implies that either $p_0 \to 0$ or $p_2 \to 0$, in which case $L_X(\bm{p}, \bm{t}) \to \infty$, as those bins lie in the support of the true distribution, but are being ruled out by the prediction.

In the asymmetric case $\alpha = 0.25$, where the high-mass bin is three times more likely to occur than the background bin, the EMD$^2$ and cross-entropy losses are minimised by $\bm{p} = (\alpha, 0, 1 - \alpha)^\top = (0.25, 0, 0.75)^\top$; thus no probability is assigned to the intermediate bin also in this more generic case. In contrast, for EMD$^1$, the expected loss is no longer constant for $\alpha \neq 0.5$: instead, it becomes linear across the simplex and is minimised by placing all weight on the more likely class. For $\alpha = 0.25$, this corresponds to $\bm{p} = (0, 0, 1)^\top$, i.e.\ a pure high-mass prediction. 
This behaviour highlights a known property of $L^1$-type losses: they encourage sparsity in the minimiser. In our case, the optimal prediction under EMD$^1$ concentrates all probability mass on the most likely class, regardless of the fact that the background occurs in 25\% of the cases. In our preliminary experiments, we found that our NN trained with the EMD$^1$ loss got stuck in a local (and unfortunately useless) minimum where the inputs were completely ignored and the NN predicted all pixels to lie in the most common class (background), suggesting that the flatter loss landscape in this case makes training harder compared to EMD$^2$, whose convexity rewards predictions that hedge between plausible outcomes.

Although this toy example only uses three bins for simplicity, our qualitative findings -- namely, that EMD$^2$ does not reward assigning probability mass to intermediate bins unsupported by the data -- are expected to generalise to more complex settings.

\section{D. Effect of primordial non-Gaussianity on Neural Network predictions}
\label{sec:app:nfl}
\begin{figure*}
    \centering
    \includegraphics[width=\linewidth]{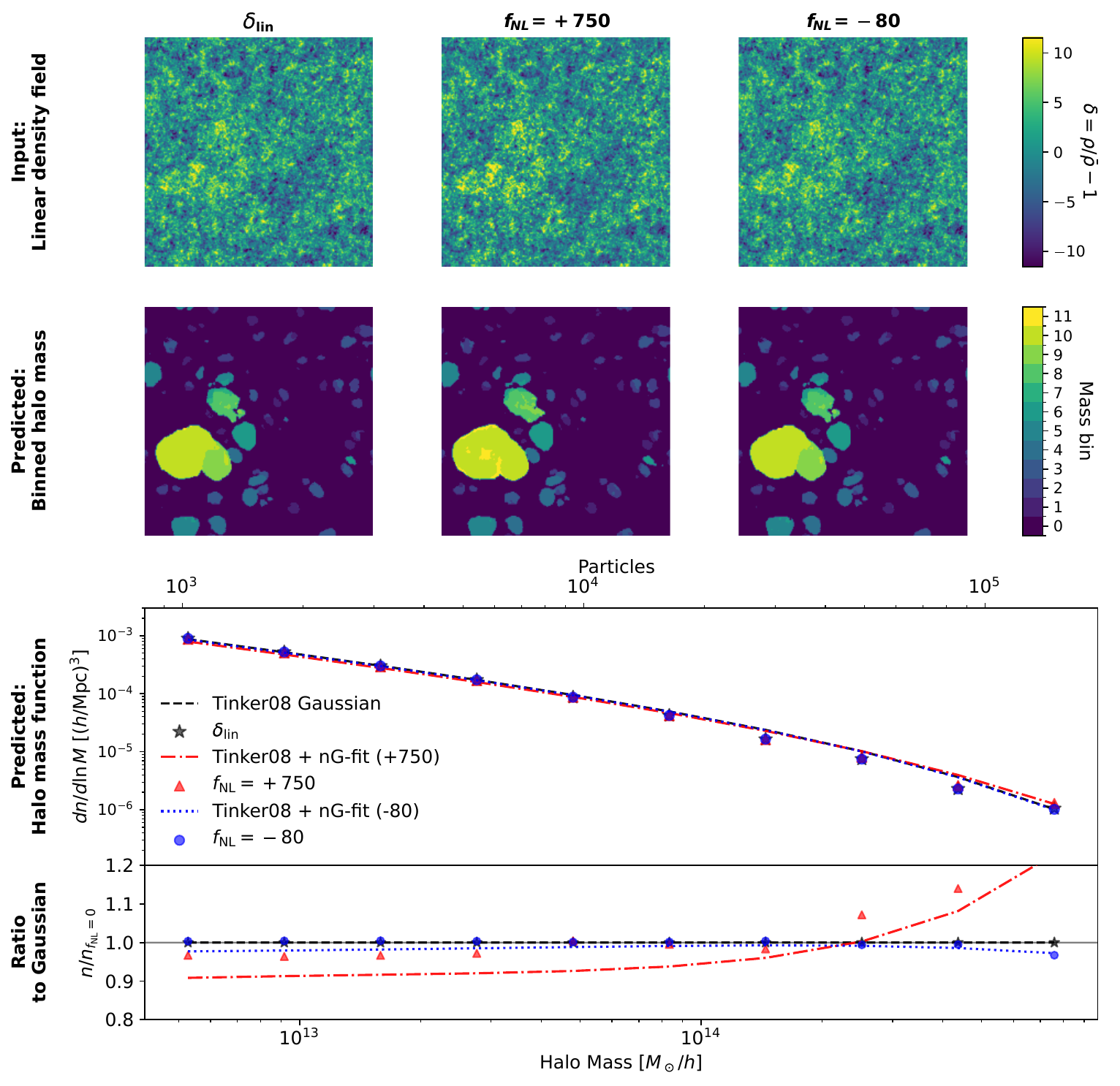}
    \caption{Comparing the effect of primordial non-Gaussianity on the NN predictions. The three different cases analysed are $f_{\text{NL}}=0$ or $\delta_{\text{lin}}$ (left), $f_{\text{NL}}=+750$ (middle), and $f_{\text{NL}}=-80$ (right). The top row shows the density fields used as input for the NN. The middle row has the predicted binned halo masses. The bottom plots show the corresponding HMF as well as the ratio of the $f_{\text{NL}}$ cases to the linear density field. The NN predictions of the HMF (scattered points) are also visually compared to analytical Tinker08 lines that are adjusted using the fitting formula from \citet{Pillepich:2010}}
    \label{fig:fnl_overview}
\end{figure*}

So far, we have only trained and predicted the NN on Gaussian initial density fields. To assess the impact of primordial non-Gaussianity on our predictions, we evaluated the model on inputs with local-type $f_{\rm{NL}} \neq 0$. As reference lines, we apply a multiplicative correction to the Tinker08 HMF,
\begin{equation} \label{eq:nfl_HMF}
\frac{dn}{d\ln M}_{\rm NG}
= \left[ \frac{f(f_{\rm NL})}{f(f_{\rm NL} = 0)} \right]
\frac{dn}{d\ln M}_{\rm T08},
\end{equation}
where $f(f_{\rm NL})$ is the HMF-correction fitting formula of \citet{Pillepich:2010}.

\autoref{fig:fnl_overview} compares the Gaussian input with the edge cases covered accurately by the fitting formula ($-80 \leq f_{\text{NL}} \leq +750 $). Even though the NN was never specifically trained on primordial non-Gaussianity, the changes in the linear density field mean that it does affect the output of the model. The difference is most clear in the middle row ($f_{\text{NL}} = +750$), which shows a strong increase in high-mass haloes. This becomes even more obvious when calculating the resulting HMF, as well as their ratios, in the bottom plot. Since we correct the analytical Tinker08 directly using the fitting formula in \autoref{eq:nfl_HMF} it already accounts for the non-Gaussianities, and we can consistently compare the NN predictions to the expected behaviour. The NN successfully reproduces the expected qualitative trend as increasing values of $f_{\text{NL}}$ lead to an enhancement of the abundance of the most massive haloes, while negative $f_{\text{NL}}$ suppresses the high-mass end of the HMF. We note that the purpose of this experiment is primarily to provide a qualitative demonstration of the NN's response when the input fields are shifted outside the training distribution. A detailed investigation of the discrepancies between the $f_{\text{NL}}$-corrected Tinker lines and the NN predictions is beyond the scope of this work. Possible contributing factors include differences in the treatment of the $\sigma_8$ normalisation after introducing the non-Gaussian component, among other effects.

\section{E. Log Derivatives}
\label{sec:app:dlog}

We already showed the dependence of the HMF on  $\Omega_m$, $n_s$ and $\sigma_8$ in \autoref{fig:dHMF-dParams_analyticals}. We now do the same, but with the logarithmic response $\partial \log \rm{HMF} / \partial \log \theta$ to reduce the dominance of large absolute abundance at low halo masses, and further connect the results to the success of \autoref{fig:MT-taylor}. We compare our NN against the same analytical models, as well as the \textsc{Mira-Titan} emulator.

\begin{figure*}
    \centering
    \includegraphics[width=\linewidth]{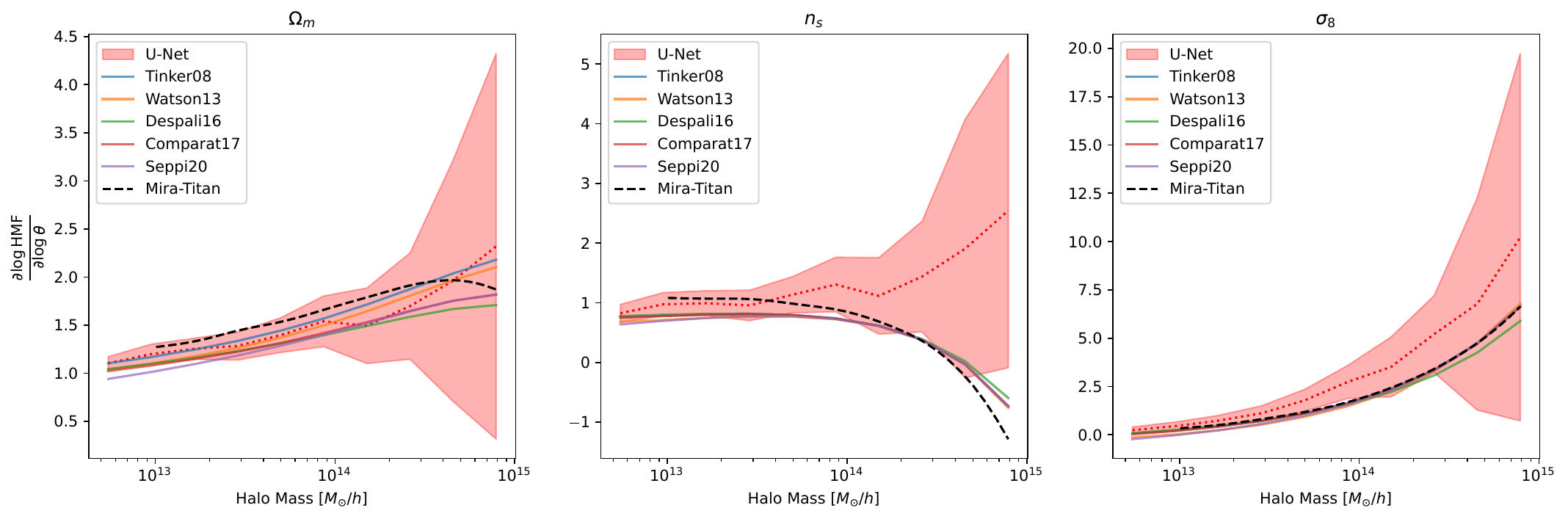}
    \caption{The logarithmic dependence of the HMF on $\Omega_m$ (left), $n_s$ (middle) and $\sigma_8$ (right), comparing our NN (red) to a variety of analytical models (solid lines) and the \textsc{Mira-Titan} emulator (black dashed). The derivatives of the U-Net are calculated over 50 different seeds to reduce cosmic variance. The resulting mean is shown as the red dashed line with the shaded red region representing the $1 \sigma$ range. The U-Net uncertainty becomes quite large for high-mass haloes, as the total abundance is so low. The largest discrepancies can be seen for $n_s$, where it fails to capture the decreasing trend, whereas the other parameter responses are reproduced much more accurately. There are still significant deviations between the models dependence, especially for $\Omega_m$.
    }
    \label{fig:dlogHMF}
\end{figure*}

\autoref{fig:dlogHMF} generally shows that the U-Net was able to learn the correct trend of the HMF on these parameters. Using the logarithmic response allows for a closer look of the high-mass regions. However, the low abundance of these haloes during training still causes the uncertainty ($1\sigma$) of our model to increase by a factor of $\approx 15$, which is why one should not put too much credibility in those areas, even though the mean is still relatively accurate. Interestingly, this representation of the dependence shows a better agreement for $\sigma_8$, but a worse one for $n_s$ when compared to \autoref{fig:dHMF-dParams_analyticals}. The mean NN prediction for $\sigma_8$ is still offset vertically, but the agreement between \textsc{Mira-Titan} and the analytical models is far smaller. Meanwhile, $n_s$ shows worse compatibility between the emulator and the other models. While the U-Net's shaded region does cover most of the theoretically computed HMF dependence, it does seem to miss the  negative correlation towards higher masses completely. $\Omega_m$ on the other hand, is predicted quite accurately as it seems. The vertical offset between these different approaches becomes even more apparent here, but our NN remains broadly consistent with the range spanned by the analytical models and is generally close to \textsc{Mira-Titan} for most mass bins. 

\vspace{5cm}

\label{lastpage}
\end{document}